\documentclass[12pt]{iopart}

\usepackage{footnote}
\expandafter\let\csname equation*\endcsname\relax
\expandafter\let\csname endequation*\endcsname\relax
\usepackage{graphics} 
\usepackage{graphicx}
\usepackage{dcolumn}
\usepackage{bm}
\usepackage{subfigure}
\usepackage{multirow}
\usepackage{float}
\usepackage{amsmath}
\usepackage{esvect}
\usepackage{amssymb}
\usepackage{pdflscape}
\usepackage[colorlinks,linkcolor=blue]{hyperref}

\begin{document}

\title{Collective expansion in pp collisions using the Tsallis statistics}
\author{Jinbiao Gu, Chenyan Li, Qiang Wang, Wenchao Zhang$^{*}$, Hua Zheng}
\address{School of Physics and Information Technology, Shaanxi Normal University, Xi'an 710119, People's Republic of China}
\eads{wenchao.zhang@snnu.edu.cn}
\begin{abstract}

  \noindent We investigate the transverse momentum ($p_{\rm T}$) spectra of identified hadrons in minimum-bias proton-proton (pp) collisions at a centre-of-mass energy ($\sqrt{s}$) of 0.9, 2.76, 5.02, 7 and 13 TeV in the framework of Tsallis-blast wave (TBW) model. It is found that the model describes well the particle spectra up to 10 GeV/c. The radial flow ($\langle \beta \rangle$) increases with the collision energy. The degrees of non-equilibrium ($q$) and the Tsallis temperature parameter ($T$) show a similar behaviour, but with a much weaker trend. With this dependence of the freeze-out parameters on the collision energy, we evaluate $\langle \beta \rangle$, $T$ and $q$ in pp collisions at $\sqrt{s}=$ 8 and 14 TeV and predict the particle spectra at these two energies. Moreover, in order to investigate the multiplicity dependence of the freeze-out parameters, the TBW model is extended to the spectra at different charged-particle multiplicity classes in pp collisions at $\sqrt{s}=$ 7 and 13 TeV. It is observed that at both energies the radial flow increases with the multiplicity while the degree of non-equilibrium shows an opposite behaviour, which is similar to that observed in proton-nucleus  (pA) and nucleus-nucleus (AA)   collisions at the Large Hadron Collider (LHC) energies. However, the Tsallis temperature parameter increases with the multiplicity, which is opposite to the trend in pA and AA collisions.  At similar multiplicities, the radial flow in pp collisions is stronger than those in pA and AA collisions, indicating that the size of the colliding system  has significant effects on the final state particle dynamics. Finally, we apply an additional flow correction to the Tsallis temperature parameter and find that the doppler-corrected temperature parameter almost scales with the multiplicity in a uniform way,  despite the difference in the colliding system and  collision energy. 
\end{abstract}

\maketitle

\section{\label{sec:intro}Introduction}
Transverse momentum ($p_{\rm T}$) spectra of identified particles are fundamental physical observables in high-energy heavy-ion collisions. They are utilized to probe different properties of the produced hot and dense matter, denoted as the quark-gluon plasma (QGP). In the low $p_{\rm T}$ region  ($p_{\rm T}\lesssim 2$ GeV/c), hadrons are produced from the soft scattering processes and the hadron spectra provide  information about the bulk system, such as the kinetic freeze-out temperature $T$ and the collective expansion velocity $\beta$.  Extraction of these properties relies on the hydrodynamic modelling of the system \cite{hydro_modelling}, such as the Boltzmann-Gibbs blast-wave (BGBW) model \cite{BGBW}.  In the high $p_{\rm T}$ region  ($p_{\rm T} \gtrsim 10$ GeV/c) , hadrons are generated by hard scatterings of partons and described by perturbative quantum chromodynamics (pQCD). 

The BGBW model has been widely used to describe the hydrodynamical expansion of the produced medium in Au-Au collisions at a centre-of-mass energy per nucleon pair ($\sqrt{s_{\rm NN}}$) of  7.7-200 GeV \cite{star_1, star_2, star_3}, in d-Au collisions at $\sqrt{s_{\rm NN}}=$ 200 GeV \cite{star_2}, in Pb-Pb collisions at $\sqrt{s_{\rm NN}}=$ 2.76 and 5.02 TeV \cite{LHC_1, LHC_2} and in p-Pb collisions at $\sqrt{s_{\rm NN}}=$ 5.02 TeV \cite{LHC_3}. In proton-proton (pp) collisions at both 7 and 13 TeV, the  $p_{\rm T}$ spectra of identified particles get harder with the increase of the charged-particle multiplicity ($dN_{\rm ch}/d\eta$), with the effect being more obvious for particles with larger mass \cite{pi_k_p_kstar_phi_vs_mult_pp_7_TeV,pi_k_p_vs_mult_pp_13_TeV}. This trend is highly similar to that observed in the evolution of the spectra in proton-nucleus (pA) and nucleus–nucleus (AA)  collisions. Thus, the BGBW model is extended to describe the pp collisions \cite{pi_k_p_kstar_phi_vs_mult_pp_7_TeV,pi_k_p_vs_mult_pp_13_TeV}. The agreement between the $p_{\rm T}$ spectra from data and the predictions from the BGBW model indicates that traces of a collective system exist in high multiplicity pp collisions. Moreover, double-ridge structures have also been observed in high multiplicity pp collisions \cite{pp_double_ridge}. These collective phenomena are reminiscent to observations attributed to the creation of QGP in Au-Au collisions at the Relativistic Heavy Ion Collider (RHIC), and in Pb-Pb collisions at the Large Hadron Collider (LHC).

In the BGBW model, it assumes that a locally thermalized medium expands with a common radial flow velocity $\beta$ and then undergoes an instantaneous kinetic freeze-out at the temperature $T$. However, in AA collisions, the fluctuation of initial conditions for hydrodynamic evolution may not be completely washed out by the subsequent interactions at either the QGP phase or the hadronic phase. In pp collisions,  from event to event there is a large energy fluctuation for particle productions. Both of the fluctuations will leave footprints on the particle spectra in the low to intermediate $p_{\rm T}$ region \cite{TBW_1}. Such a situation in AA collisions\cite{Tsallis_8,Tsallis_9,Tsallis_10,Tsallis_11,Tsallis_12} and in pp collisions \cite{Tsallis_1,Tsallis_2,Tsallis_3,Tsallis_4,Tsallis_5,Tsallis_6,Tsallis_7}  is coped with a non-extensive  statistics, i.e., the Tsallis statistics \cite{Tsallis_0}. In order to take both the fluctuation and the collective expansion into account, the Tsallis statistics is embedded into the BGBW model. The Tsallis blast-wave (TBW) model has been utilized to describe the identified particle spectra in Pb-Pb, Xe-Xe and p-Pb collisions at the LHC energies \cite{TBW_2} and in Au-Au collisions at the RHIC energies \cite{TBW_1, TBW_3, TBW_4}. In ref. \cite{TBW_5}, this model is extended to describe the spectra in pp collisions at a centre-of-mass energy ($\sqrt{s}$) of 0.2, 0.54, 0.9 and 7 TeV. The authors found an onset of radial flow in pp collisions when the collision energy is above 0.9 TeV.

In this paper, as a complementary study to that performed in ref. \cite{TBW_5}, the TBW model is applied to describe the identified particle $p_{\rm T}$ spectra in minimum-bias\footnote{In order to distinguish the spectra with and without the multiplicity cut in pp collisions, the phrase “minimum-bias'' is introduced for the latter.} pp collisions at $\sqrt{s}=$ 0.9, 2.76, 5.02, 7 and 13 TeV. This study can provide the collision energy dependence of the radial flow velocity, the Tsallis temperature parameter as well as the degree of non-equilibrium in small systems. With this dependence, we predict the identified particle spectra in pp collisions at $\sqrt{s}=$ 8 and 14 TeV. Moreover,  in order to investigate the multiplicity dependence of the freeze-out parameters, the TBW model is extended to the spectra at different charged-particle multiplicity classes in pp collisions at $\sqrt{s}=$ 7 and 13 TeV. Combined with our previous results in Pb-Pb, Pb-Pb, Xe-Xe and p-Pb collisions at $\sqrt{s_{\rm NN}}=$ 2.76, 5.02, 5.44 and 5.02 TeV \cite{TBW_2}, such a systematic study on the multiplicity dependence of freeze-out parameters for different colliding systems at the LHC may shed some light on the possible underlying mechanism for the particle production.

The paper is organized as follows. In sect. \ref{sec:method}, we briefly introduce the TBW model. In sect. \ref{sec:results}, results and discussions are presented. Finally, the conclusion is given in sect. \ref{sec:conclusions}.

\section{\label{sec:method}The TBW model}
With the recipe of the TBW model in refs. \cite{TBW_1, TBW_2,TBW_3, TBW_4,TBW_5}, the invariant differential yield of identified particles at mid-rapidity in pp collisions is expressed as\footnote{Considering the rapidity distribution of the emitting source, we introduce an extra exponential term, $dN/dy_s=\textrm{exp}(\sqrt{y_b^2-y_s^2})$, in the model presented in ref. \cite{TBW_2}.}
\begin{eqnarray}\label{eq:TBW}
\frac{d^{2} N}{2 \pi p_{\rm T} d p_{\rm T} d y} \propto &&m_{T} \int_{-y_b}^{+y_b} e^{\sqrt{y_b^2-y_s^2}}\cosh \left(y_{s}\right) d y_{s}\int_{-\pi}^{+\pi} d \phi_{}  \nonumber \\
	&& \times \int_{0}^{R} rd r \bigg[1+\frac{q-1}{T}\Big(m_{\rm T} \cosh \left(y_{s}\right) \cosh \left(\rho \right)\nonumber \\
	&& - p_{\rm T} \sinh \left(\rho \right) \cos \left(\phi \right)\Big)\bigg]^{-1 /(q-1)},
\end{eqnarray}
where
\begin{equation}
	m_{\rm T}=\sqrt{p_{\rm T}^2+m^2},
\end{equation}
\begin{equation}
	y_b=\ln(\sqrt{s}/m_{\rm N}),
\end{equation}
\begin{equation}
	\rho=\tanh^{-1}\left[\beta_s\left(\frac{r}{R}\right)^n\right].
\end{equation}
 $y$, $y_{s}$ and $y_b$ are, respectively, the rapidities of the produced particle, the emitting source and the beam; $m$ ($m_{\rm N}$) is the mass of the produced particle (the colliding nucleon); $\phi$ is the difference between the azimuthal angles of the emitted particle velocity ($\phi_{p}$) and the boost of the source element ($\phi_{b}$) with respect to the reaction plane.  $\phi_{b}$ is deemed as the same as the azimuthal angle of the source element in coordinate space, $\phi_{s}$.  $T$ is the Tsallis temperature parameter, $q$ is a non-extensive parameter which describes the deviation of the system from thermal equilibrium. As described in ref. \cite{q_meaning}, the Tsallis distribution is interpreted as a superposition of Boltzmann-Gibbs distributions with different temperatures. The fluctuation of these temperatures is given by the deviation of $q$ from unity, while the average value of their reciprocals represents $1/T$. When $q\rightarrow 1$, one recovers the distribution in the BGBW model.  The transverse flow rapidity, $\rho$, grows as the $n$th power of the radial distance ($r$) in the transverse plane and $\beta_{s}$ is the transverse flow velocity at the surface of the fireball ($r=R$). The average transverse flow velocity is expressed as $\langle \beta \rangle =2/(n+2)\beta_{s}$. In the model, the default value of $n$ is set as 1 and $\langle \beta \rangle =2/3\beta_{s}$.  In our recent paper, we found that in central AA collisions the freeze-out parameters rely on the choice of $n$ \cite{TBW_2}. In this paper, in order to investigate whether a similar dependence emerges in pp collisions, besides the linear velocity profile, a constant velocity profile with $n=0$\footnote{For the constant profile, $\langle \beta \rangle =\beta_{s}$.} is also considered. This profile was first applied to the description of the identified particle's elliptic flow in Au-Au collisions with the BGBW model \cite{TBW_v2}. For both transverse velocity profiles, four free parameters are to be extracted: the normalization constant, $\langle \beta \rangle$, $q$ and $T$. 

\section{\label{sec:results}Results and discussions}
The ALICE collaboration has presented a series of identified particle spectra in pp collisions at $\sqrt{s}=$ 0.9\footnote{For pp collisions at $\sqrt{s}= $ 0.9 TeV, the spectra data of $K^{\star0}$ and $\Omega$ are not yet available.}, 2.76, 5.02\footnote{The spectra of strange particles in pp collisions at $\sqrt{s}= $ 5.02 TeV were published by the CMS collaboration.}, 7 and 13 TeV \cite{pi_k_p_pp_0_9_TeV,ks_phi_lambda_xi_pp_0_9_TeV,pi_k_p_pp_2_76_TeV,Kstar_phi_pp_2_76_TeV,xi_omega_pp_2_76_TeV,lambda_pp_2_76_TeV,pi_k_p_pp_5_02_TeV,kstar_phi_pp_5_02_TeV,ks_lambda_xi_omega_pp_5_02_TeV,kstar_phi_pp_7TeV,ks_lambda_in_pp_7TeV_and_all_in_pp_13TeV,Omega_Xi_pp_7TeV}. Recently, they also published the spectra with different charged-particle multiplicity classes in pp collisions at $\sqrt{s}=$ 7 and 13 TeV \cite{pi_k_p_kstar_phi_vs_mult_pp_7_TeV,pi_k_p_vs_mult_pp_13_TeV,strange_vs_mult_pp_7TeV,kstar_phi_vs_mult_pp_13TeV,strange_mult_pp_13_TeV}. Table \ref{tab:dataRef} shows the references for the spectra of identified particles at a given collision energy. Here, $\pi$, $K$, $p$, $K^{*0}$, $\Lambda$, $\Xi$ and $\Omega$, respectively, refer to $\pi^{+}+\pi^{-}$, $K^{+}+K^{-}$, $p+\bar{p}$, $K^{*0}+\bar{K}^{*0}$, $\Lambda+\bar{\Lambda}$, $\Xi^{-}+\bar{\Xi}^{+}$ and $\Omega^{-}+\bar{\Omega}^{+}$.

\begin{table}[H]
  \caption{Summary of references for the identified particle spectra at a given collision energy.}\label{tab:dataRef}
  \begin{center}
	\begin{tabular}{l@{\hspace{0.5cm}}c@{\hspace{0.5cm}}c@{\hspace{0.5cm}}c}
		\hline
		&$\pi$, $K$, $p$ & $K^{\star0}$, $\phi$   &   $K_S^0$, $\Lambda$, $\Xi$, $\Omega$\\
		\hline
		0.9 TeV  & \cite{pi_k_p_pp_0_9_TeV} & \cite{ks_phi_lambda_xi_pp_0_9_TeV} & \cite{ks_phi_lambda_xi_pp_0_9_TeV} \\
		2.76 TeV & \cite{pi_k_p_pp_2_76_TeV} & \cite{Kstar_phi_pp_2_76_TeV} & \cite{xi_omega_pp_2_76_TeV,lambda_pp_2_76_TeV} \\
		5.02 TeV & \cite{pi_k_p_pp_5_02_TeV} &     \cite{kstar_phi_pp_5_02_TeV}    & \cite{ks_lambda_xi_omega_pp_5_02_TeV}    \\
		7 TeV  & \cite{pi_k_p_kstar_phi_vs_mult_pp_7_TeV,pi_k_p_pp_5_02_TeV}  & \cite{pi_k_p_kstar_phi_vs_mult_pp_7_TeV,kstar_phi_pp_7TeV}  &  \cite{pi_k_p_kstar_phi_vs_mult_pp_7_TeV,ks_lambda_in_pp_7TeV_and_all_in_pp_13TeV,Omega_Xi_pp_7TeV,strange_vs_mult_pp_7TeV}     \\
		13 TeV & \cite{pi_k_p_vs_mult_pp_13_TeV,ks_lambda_in_pp_7TeV_and_all_in_pp_13TeV}   & \cite{ks_lambda_in_pp_7TeV_and_all_in_pp_13TeV,kstar_phi_vs_mult_pp_13TeV}   & \cite{ks_lambda_in_pp_7TeV_and_all_in_pp_13TeV,strange_mult_pp_13_TeV} \\
		\hline
	\end{tabular}
        \end{center}
\end{table}



\begin{landscape}
\begin{figure}[H]
 \centering
\includegraphics[scale=0.32]{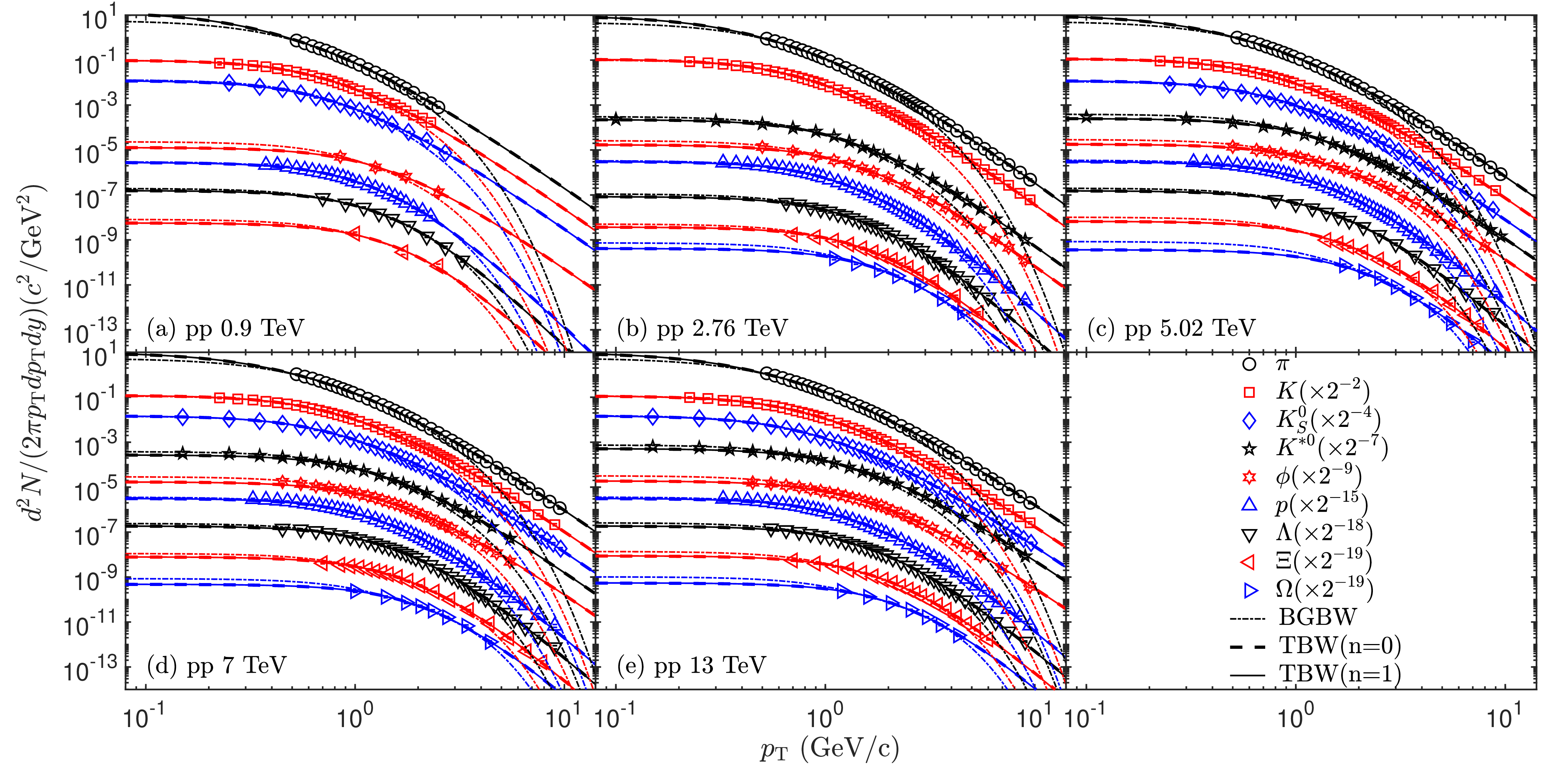}
\caption{\label{fig:pt_spectra} (Colour online) Identified particle $p_{\rm T}$ spectra in pp collisions at $\sqrt{s}=$ 0.9 TeV (a), 2.76 TeV (b), 5.02 TeV (c), 7 TeV (d) and 13 TeV (e). The symbols are experimental data taken from refs. \cite{pi_k_p_pp_0_9_TeV,ks_phi_lambda_xi_pp_0_9_TeV,pi_k_p_pp_2_76_TeV,Kstar_phi_pp_2_76_TeV,xi_omega_pp_2_76_TeV,lambda_pp_2_76_TeV,pi_k_p_pp_5_02_TeV,kstar_phi_pp_5_02_TeV,ks_lambda_xi_omega_pp_5_02_TeV,kstar_phi_pp_7TeV,ks_lambda_in_pp_7TeV_and_all_in_pp_13TeV,Omega_Xi_pp_7TeV}. The solid (dashed) curves represent the results from the TBW model with the linear (constant) velocity profile, while the dash-dotted lines are the results from the BGBW model.}
\end{figure}
\end{landscape}

\begin{landscape}
\begin{figure}[H]
  \centering
\includegraphics[scale=0.32]{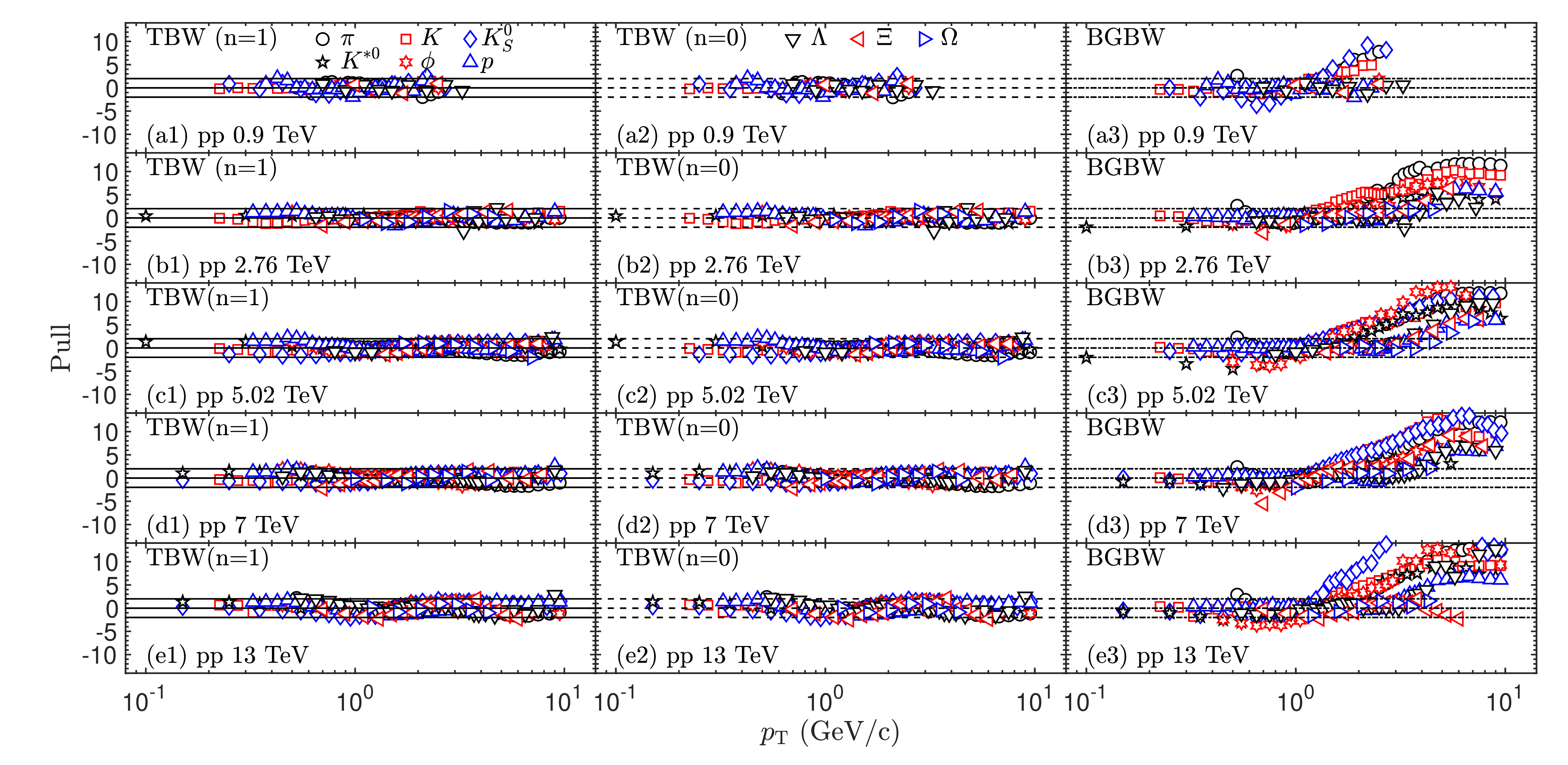}
\caption{\label{fig:pull_dist} (Colour online) The pull distributions for the TBW model with $n=1$ (left column), $n=0$ (middle column) and the BGBW model (right column) in pp collisions at $\sqrt{s}=$ 0.9, 2.76, 5.02, 7 and 13 TeV from top to bottom panels. In a given subfigure, the upper and lower solid (dashed or dash-dotted) lines represent that the deviation between the model and the data is 2 times of the data's error.}
\end{figure}
\end{landscape}

In order to investigate the evolution of the freeze-out parameters as a function of $\sqrt{s}$, a combined fit is performed to the identified particle spectra with the TBW model in minimum-bias pp collisions at a given collision energy using a least $\chi^{2}$ method. At low $p_{\rm T}$, for pions there is a large feed-down contribution from resonance decays, which will steepen the spectra \cite{LHC_1}. To remove this contribution, the pion spectrum at the low $p_{\rm T}$ region ($p_{\rm T}<0.5$ GeV/c) is excluded from the fit. This cut has been widely used in the BGBW parameterization of the particle spectra in AA and pp collisions by experimental groups \cite{star_1, star_2, star_3,LHC_1, LHC_2, LHC_3,pi_k_p_kstar_phi_vs_mult_pp_7_TeV,pi_k_p_vs_mult_pp_13_TeV}. For other particles, we do not apply any lower $p_{\rm T}$ cut on their spectra. In the fit, the average flow velocity $\langle \beta \rangle$ and the Tsallis temperature parameter $T$ are common for all hadrons, while the degrees of non-equilibrium are, respectively, universal for the meson group ($q_M$) and the baryon group ($q_B$). It is found that the separation of mesons and baryons shows good fits while one single $q$ for all particles gives a bad description of the spectra. In addition, different particles own different normalization factors and masses. The freeze-out parameters of the combined TBW fit in pp collisions at the given collision energy are tabulated in table \ref{tab:parameter_0_9_13_TeV}.  Also shown in the table are the $\chi^2$ values divided by the number of degrees of freedom ($\chi^2$/NDF). The first uncertainty in the table is the statistical error returned from the combined fit. The second one is the systematic uncertainty which is due to the variation of the lower bound (from 0.5 to 0.1 GeV/c) for the pion spectrum. Fig. \ref{fig:pt_spectra}(a) ((b), (c), (d) and (e)) presents the identified particle spectra with their associated TBW model results in pp collisions at $\sqrt{s}=$ 0.9 (2.76, 5.02, 7 and  13) TeV. At each collision energy, in a logarithmic scale the curves from the model with both $n=1$ (solid lines) and $n=0$ (dashed lines) describe well the data  over a broad $p_{\rm T}$ range (0-10 GeV/c). In order to quantify the agreement between the data points and the fitted curves, a variable pull=(data-fitted)/$\rm \Delta data$\footnote{$\Delta$data is the square root of the quadratic sum of the data’s statistical and systematic uncertainties.} is evaluated. The pull distributions in the left and middle columns of Fig. \ref{fig:pull_dist} show that most of the data are consistent with the TBW model within 2 standard deviations. As a comparison, the BGBW calculations (dash-dotted lines) are also presented in Fig. \ref{fig:pt_spectra}. As the average transverse flow velocity and the kinetic freeze-out temperature of the BGBW model for minimum-bias pp collisions are not available so far in literature, they are determined by fitting the model simultaneously to the spectra of charged pions, kaons and protons in the ranges 0.5-1 GeV/c, 0.2-1.5 GeV/c and 0.3-3 GeV/c, respectively. The choice of particle species as well as the BGBW fit range is consistent with that from previous publications of the ALICE collaboration \cite{pi_k_p_kstar_phi_vs_mult_pp_7_TeV,pi_k_p_vs_mult_pp_13_TeV}. As it can be seen from Fig.  \ref{fig:pull_dist}, due to the lack of the knob for the non-equilibrium degree, the BGBW model tends to underestimate the spectra for $\pi$, $K$, $K_S^{0}$, $K^{*0}$, $\phi$, $p$, $\Lambda$, $\Xi$ and $\Omega$ when $p_{\rm T}$ is, respectively, greater than 1 GeV/c, 1.5 GeV/c,  1.5 GeV/c,  2 GeV/c, 1.5 GeV/c, 3 GeV/c, 3 GeV/c, 4 GeV/c and  4 GeV/c. For $\phi$ and $\Xi$, deviations also appear for $p_{\rm T}<$ 1 GeV/c.

\begin{figure}[H]
   \centering
\includegraphics[scale=0.26]{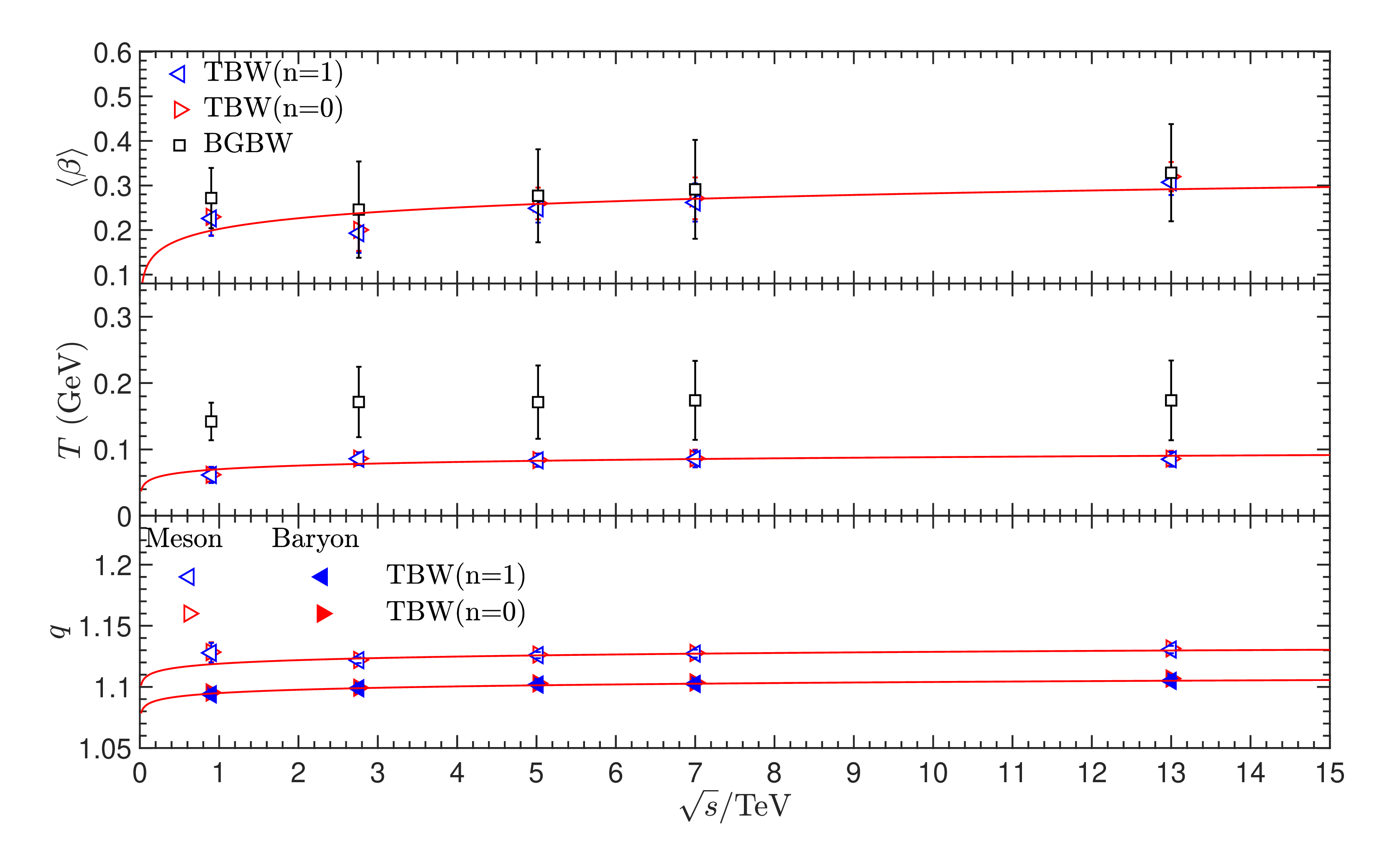}
\caption{\label{fig:beta_T_q_vs_energy} (Colour online) $\langle \beta \rangle$, $T$, $q_{M}$ and $q_{B}$ as a function $\sqrt{s}/\rm TeV$. The solid curves represent a logarithmic fit for the dependence of the freeze-out parameters on the collision energy. The error bars in the upper, middle and lower panels, respectively, represent the total uncertainties of $\langle \beta \rangle$, $T$, $q_{M}$ and $q_{B}$.}
\end{figure}

With the parameters in table  \ref{tab:parameter_0_9_13_TeV}, we present the dependence of $\langle \beta \rangle$, $T$, $q_{B}$ and $q_{M}$ for the TBW model on the collision energy in Fig. \ref{fig:beta_T_q_vs_energy}. Also shown in the figure are the freeze-out parameters for the BGBW model. The solid curves represent  logarithmic parameterizations of the dependence. Some results can be obtained from the figure as follows.

(i)  The freeze-out parameters from the TBW model with $n=1$ are compatible with those from the model with $n=0$ within uncertainties. This is similar to that observed in peripheral Pb–Pb (Pb-Pb, Xe–Xe, p–Pb) collisions at $\sqrt{s_{\rm NN}}=$ 2.76 (5.02, 5.44, 5.02) TeV\cite{TBW_2}.

(ii) For the TBW model, the average transverse flow velocity $\langle \beta \rangle$ grows with $\sqrt{s}$.  The growth can be parameterized as $\langle \beta \rangle=(0.202\pm0.026)+(0.035\pm0.014)\ln(\sqrt{s}/\rm TeV)$.  It indicates that the system is more explosive at higher energies. Similar trend was observed in minimum-bias pp collisions at lower energies \cite{TBW_5} and in central Au-Au collisions at the RHIC energies \cite{star_3,TBW_3}. In pp collisions at $\sqrt{s}=$ 13 TeV, $\langle \beta \rangle$ reaches an average value of $0.307\pm 0.007 \pm 0.028$, which is comparable with that in peripheral (60-80$\%$) Pb-Pb collisions at $\sqrt{s_{\rm NN}}=$ 2.76 TeV, $0.293 \pm 0.012 \pm 0.014$ \cite{TBW_2}. 

(iii) Compared to $\langle \beta \rangle$, the Tsallis temperature parameter $T$ shows a similar but much weaker dependence on $\sqrt{s}$, which is described by  $T=(0.070\pm0.006)+(0.008\pm0.004)\ln(\sqrt{s}/\rm TeV)$. This dependence is similar to that observed in refs. \cite{T_dependence_1,T_dependence_2}, in which the authors argued that at higher collision energies the system has a higher excitation. However, in some literature the authors claimed that the kinetic freeze-out temperature showed decreasing trend \cite{star_3,T_dependence_3} or little dependence with the increase of the collision energy \cite{T_dependence_5,T_dependence_4}. 
 

(iv) Both of the non-extensive parameters $q_{M}$ and $q_{B}$ show a weak increase with $\sqrt{s}$. This trend is similar to that observed in pp collisions at lower collision energy \cite{TBW_5} and in peripheral  Au-Au collisions \cite{TBW_3}. The increase can be separately described as $q_{M}=(1.1190\pm0.0028)+(0.0042\pm0.0016)\ln(\sqrt{s}/\rm TeV)$ and $q_{B}=(1.0949\pm0.0005)+(0.0039\pm0.0003)\ln(\sqrt{s}/\rm TeV)$. It suggests that the system is more out of thermal equilibrium at higher collision energy \cite{TBW_3}.


(v) At a given $\sqrt{s}$, $q_{M}$ is larger than $q_{B}$. Similar conclusions are obtained in pp collisions at lower collision energy \cite{TBW_5}. This could be explained as follows. At high $p_{\rm T}$, hard process dominates the particle production. It has been shown  in refs. \cite{power_law_2,power_law_3} that the parton-parton hard scattering in pp collisions leads to the hadron $p_{\rm T}$ distribution that resembles the Tsallis distribution. Moreover, the hadron spectra from hard scattering behave as a power law distribution, $p_{\rm T}^{-n}$. The index $n$ is related to the non-extensive parameter $q$ with $n=1/(q-1)$.  As shown in refs. \cite{power_law_2,power_law_3,power_law_1}, $n$ is expressed as $n=2n_{a}-4$, where $n_a$ is the number of active participating quarks. If the dominant processes for the hadron production are parton-parton hard scatterings $qq\rightarrow qq$ (referred as the \textit{leading twist process}), then the counting rule will give $n=4$. The index $n$ will become larger if the contribution from higher twist processes is taken into account. For example, for the meson scattering process $q+\textrm{meson}\rightarrow q+\textrm{meson}$, $n=8$, while for the baryon scattering process $q+\textrm{baryon}\rightarrow q+\textrm{baryon}$,  $n=12$.  This naturally gives $q_{M}>q_{B}$.

(vi) The average transverse flow velocity $\langle \beta \rangle$ is larger in the BGBW model  than in the TBW model. Moreover, the kinetic freeze-out temperature in the former is higher than the Tsallis temperature parameter in the latter. This can be understood as follows. At fixed values of $T$ and $\langle \beta \rangle$, the TBW distribution with $q>1$ is always larger than the BGBW distribution. As a consequence, in order to keep the particle yields the same, the BGBW distribution has to boost its radial flow parameter as well as the kinetic freeze-out temperature for the same set of particle yields.



\begin{landscape}
\begin{figure}[H]
   \centering
\includegraphics[scale=0.32]{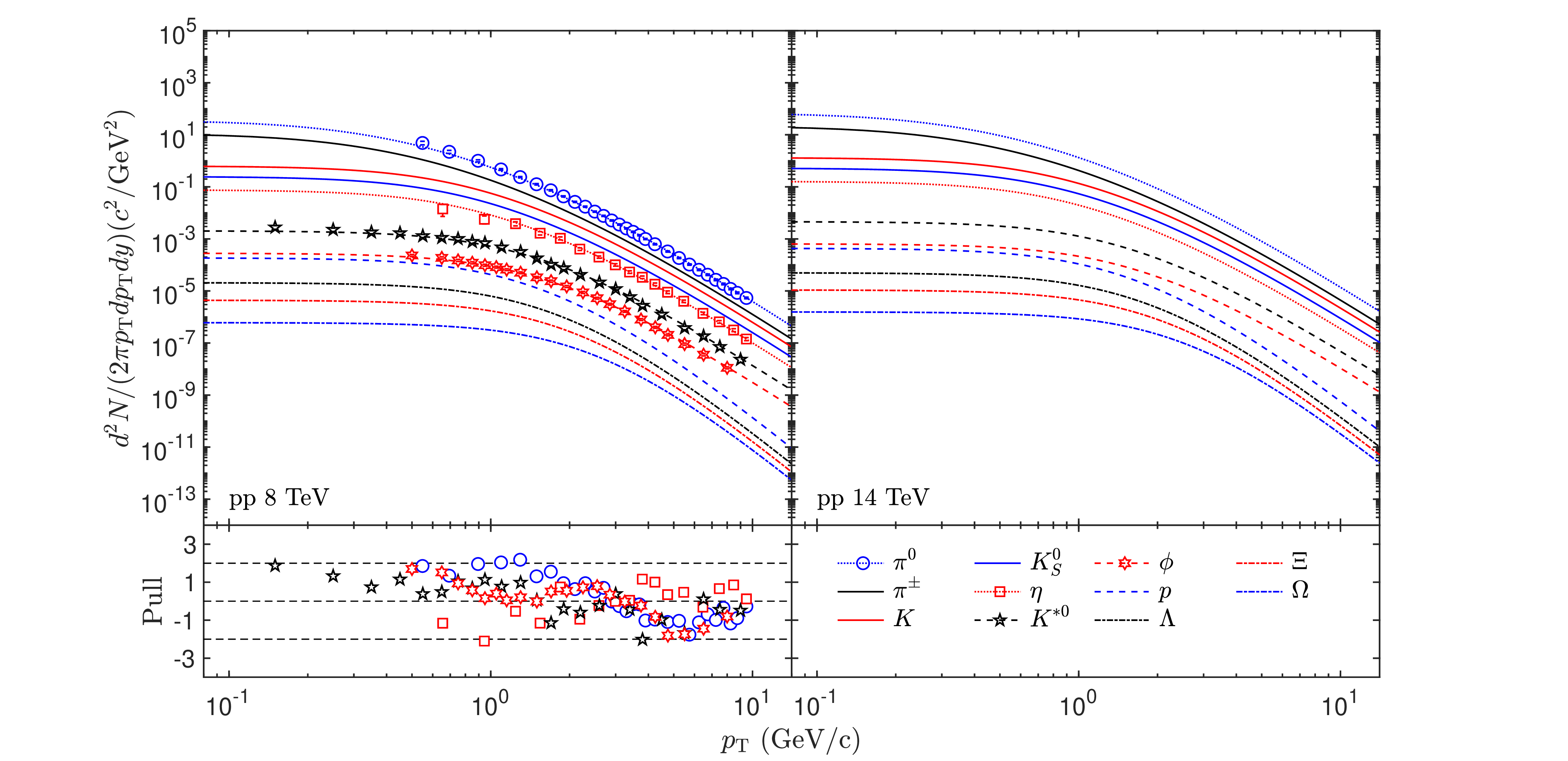}
\caption{\label{fig:predicted_8_14_TeV} (Colour online) Identified particle $p_{\rm T}$ spectra in pp collisions at $\sqrt{s}=$ 8 TeV (the upper left panel) and 14 TeV (the upper right panel). The solid, dashed, dash-dotted and dotted curves represent the spectra predicted from the TBW model. The different symbols in the left panel are the experimental spectra of $\pi^0$, $\eta$, $K^{*0}$ and $\phi$ \cite{kstar_phi_pp_7TeV,pi0_eta_8_TeV}. In the lower left panel, the upper and lower dashed lines represent that the deviation between the prediction and the data is 2 times of the data's error. }
\end{figure}
\end{landscape}

With the parameterization of the freeze-out parameters on the collision energy, we could evaluate $\langle \beta \rangle$, $T$ and $q$ in pp collisions at $\sqrt{s}=$ 8 TeV and 14 TeV. For the former, $\langle \beta \rangle$=$0.275\pm0.055$, $T=0.087\pm0.014$ GeV, $q_M=1.1278\pm 0.0061$ and $q_B=1.1031\pm 0.0011$, while for the later, $\langle \beta \rangle$=$0.294\pm0.062$, $T=0.091\pm0.016$ GeV, $q_M=1.1301\pm 0.0069$ and $q_B=1.1054\pm 0.0013$. With these values, we can predict the spectra of identified particles in pp collisions at these two collision energies. The solid, dashed, dash-dotted and dotted curves in the upper panels of Fig. \ref{fig:predicted_8_14_TeV} represent the predicted spectra. Also shown in the figure are the experimental spectra of $\pi^0$, $\eta$, $K^{*0}$ and $\phi$ in pp collisions at 8 TeV \cite{kstar_phi_pp_7TeV,pi0_eta_8_TeV}.  It is observed that the predicted curves describe well the data points of these four identified particles. From the pull distribution in the lower left panel of the figure, we see that these data are consistent with the TBW predictions within 2 standard deviations. The corresponding $\chi^2$/NDF value for the prediction is 1.04. This implies that our prediction of the $p_{\rm T}$ spectra from the TBW model is reliable.

\begin{landscape}
\begin{figure}[H]
  \centering
\includegraphics[scale=0.32]{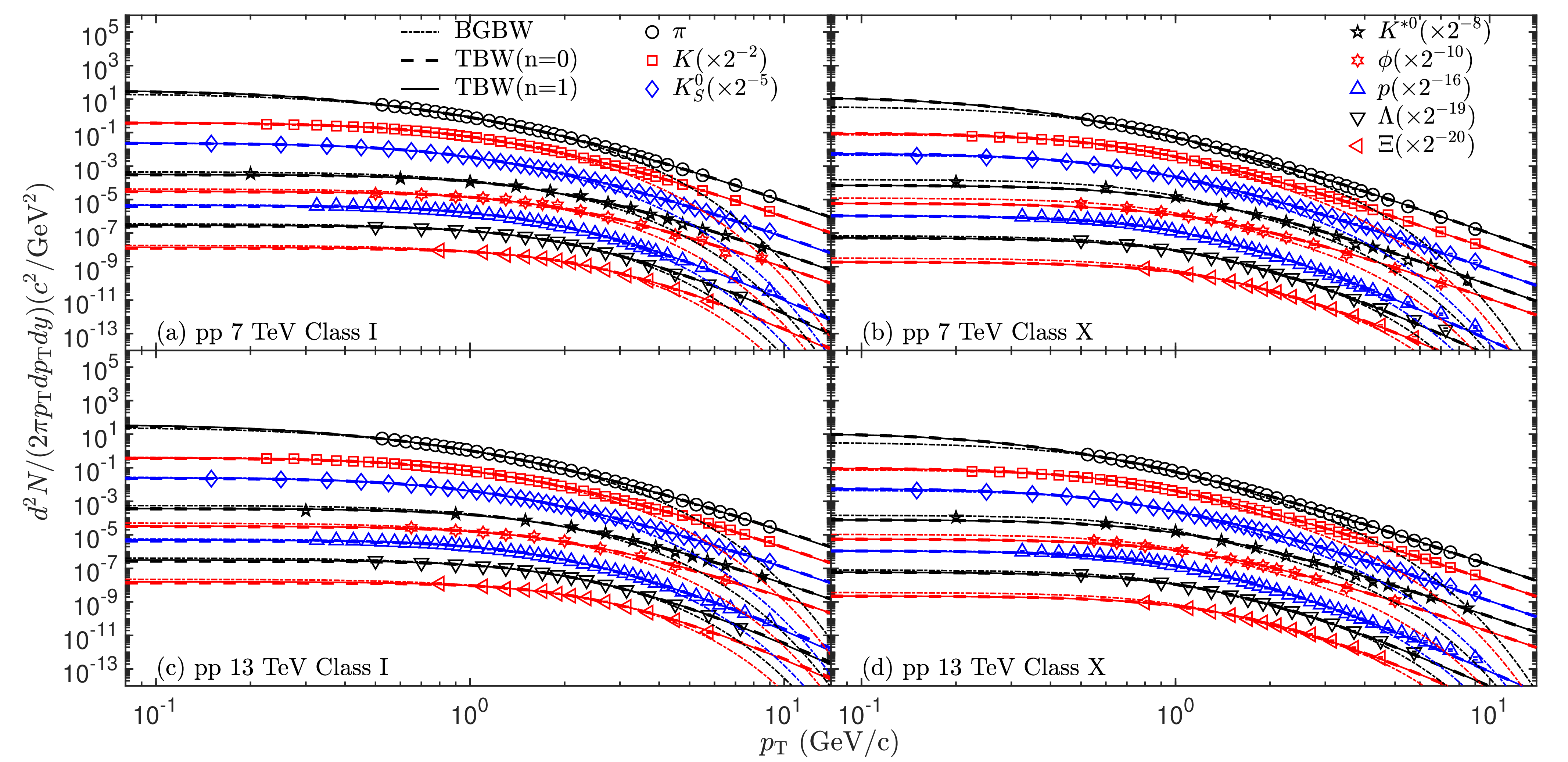}
\caption{\label{fig:pp_7_and_13_class_I_X_TeV_spectra} (Colour online) Identified particle $p_{\rm T}$ spectra at the multiplicity class I (left column) and class X (right column) in pp collisions at $\sqrt{s}=$ 7 and 13 TeV from top to bottom panels. The symbols are experimental data  taken from refs. \cite{pi_k_p_kstar_phi_vs_mult_pp_7_TeV,strange_vs_mult_pp_7TeV, pi_k_p_vs_mult_pp_13_TeV,kstar_phi_vs_mult_pp_13TeV,strange_mult_pp_13_TeV}. The solid (dashed) curves represent the results from the TBW model with the linear (constant) velocity profile, while the dash-dotted lines are the results from the BGBW model. }
\end{figure}
\end{landscape}

\begin{landscape}
\begin{figure}[H]
  \centering
\includegraphics[scale=0.32]{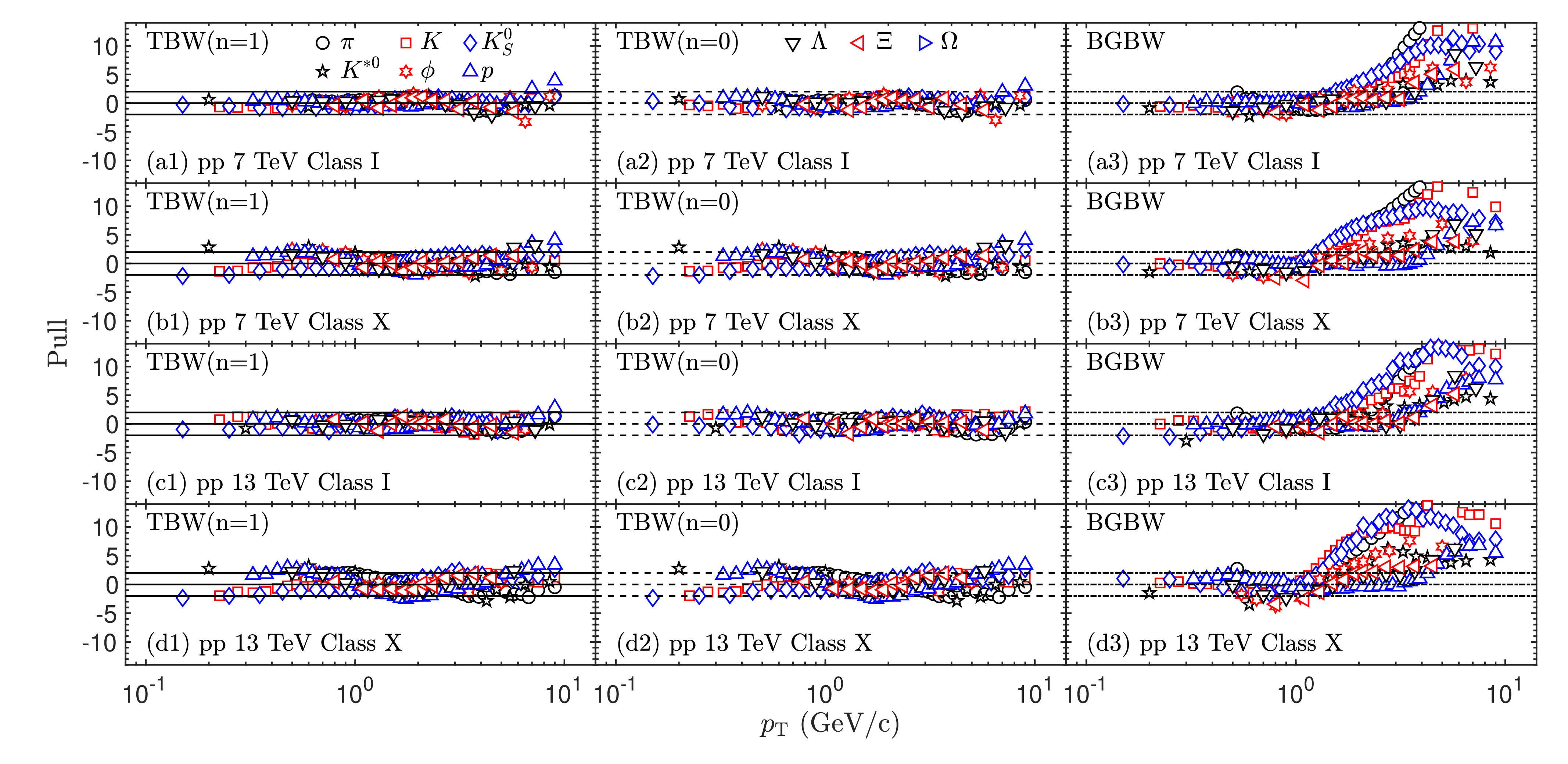}
\caption{\label{fig:pp_7_and_13_class_I_X_TeV_pull} (Colour online) The pull distributions for the TBW model with $n=1$ (left column), $n=0$ (middle column) and the BGBW model (right column) at the multiplicity classes I and X in pp collisions at $\sqrt{s}=$ 7 and 13 TeV from top to bottom panels. In a given subfigure, the upper and lower solid (dashed or dash-dotted) lines represent that the deviation between the model and the data is 2 times of the data's error. }
\end{figure}
\end{landscape}

\begin{figure}[H]
  \centering
\includegraphics[scale=0.20]{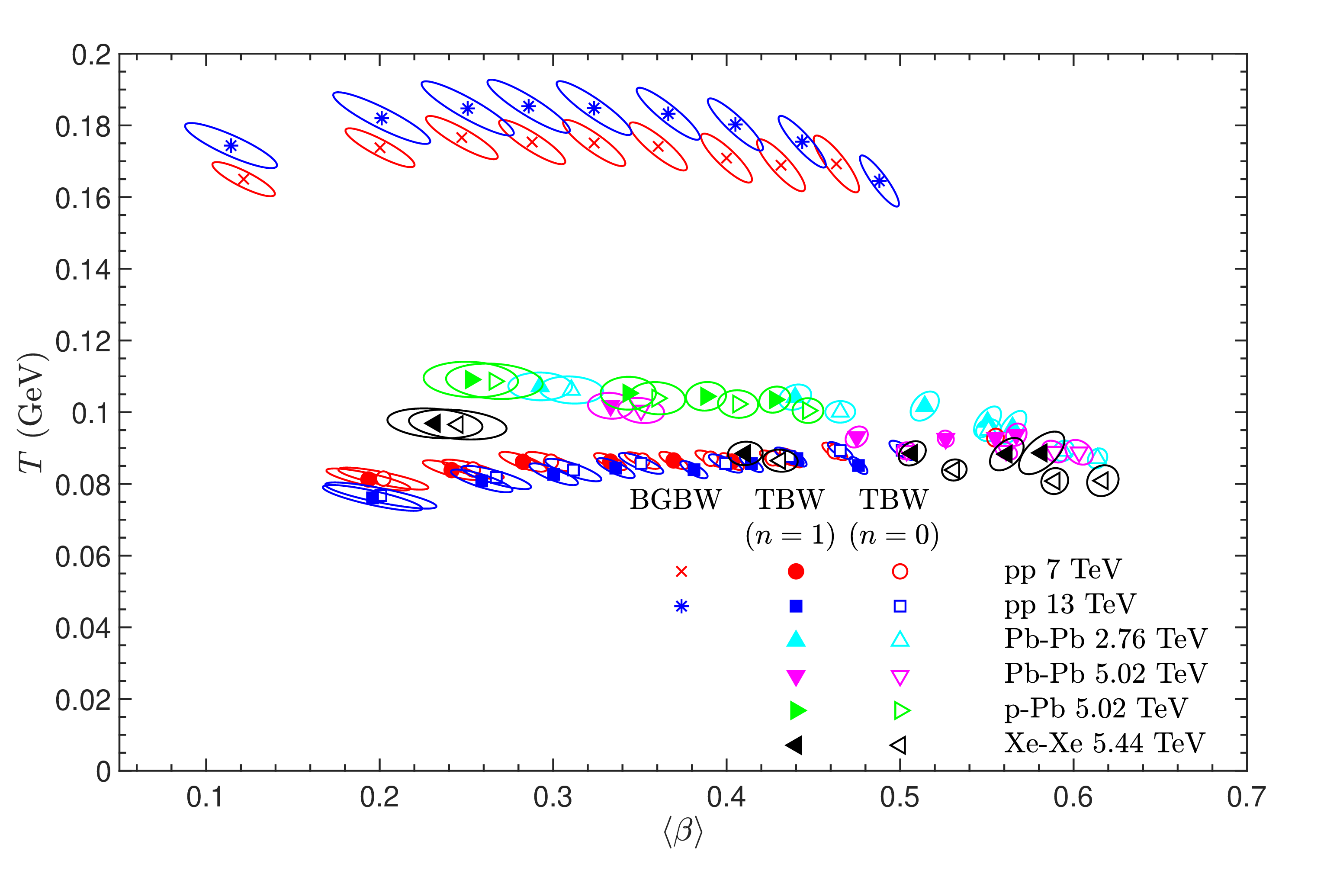}
\caption{\label{fig:beta_T} (Colour online)  Correlation of $T$ and $\langle \beta \rangle$ for the TBW model in pp (Pb-Pb, Pb-Pb, Xe-Xe, p-Pb) collisions at $\sqrt{s}=$ 7 and 13 ($\sqrt{s_{\rm NN}}=$ 2.76, 5.02, 5.44, 5.02) TeV. Also shown in the figure are the correlations between $T$ and $\langle \beta \rangle$ for the BGBW model in pp collisions at $\sqrt{s}=$ 7 and 13 TeV. The elliptic contours reflect 1 $\sigma$ uncertainty. The charged-particle multiplicity increases from left to right for a given colliding system at a selected collision energy.}
\end{figure}

In order to investigate the multiplicity dependence of the freeze-out parameters, we extend the investigation to the identified particle spectra with different charged-particle multiplicity classes in pp collisions at $\sqrt{s}=$ 7  and 13 TeV. The values of $\langle dN_{\rm ch}/d\eta \rangle$ at different multiplicity classes in pp collisions can be found in table \ref{tab:pp_diff_mult}\footnote{The values of $\langle dN_{\rm ch}/d\eta \rangle$ at different centralities in pA and AA collisions can be found in table \ref{tab:pA_AA_diff_mult}.}.  The multiplicity classe selection is based on the cuts on the total charge deposited in the V0 detectors (V0M amplitude), as the V0M amplitude scales linearly with the total number of the corresponding charged particles in the detector’s acceptance. For the spectra at the multiplicity classes IX and X (class X)  in pp collisions at $\sqrt{s}=$ 7 (13) TeV, $\langle \beta \rangle$ returned from the combined fit is as low as $10^{-5}$. Thus we fix it as 0 and repeat the analyses. Figure  \ref{fig:pp_7_and_13_class_I_X_TeV_spectra} shows the identified particle spectra together with the TBW results at two selected multiplicity classes (classes I and X) for these two collision energies. At each collision energy, the TBW model generally reproduces the data at both multiplicity classes well. Also shown in the figure are the BGBW results with $\langle \beta \rangle$ and $T$ taken from refs. \cite{pi_k_p_kstar_phi_vs_mult_pp_7_TeV,pi_k_p_vs_mult_pp_13_TeV}. As can be seen from Fig. \ref{fig:pp_7_and_13_class_I_X_TeV_pull}, the agreement between the data and the TBW model results is within 2 standard deviations. However,  for the BGBW model, it underestimates the data of $\pi$, $K$, $K_S^{0}$, $K^{*0}$, $\phi$, $p$, $\Lambda$ and  $\Xi$ when $p_{\rm T}$ is, respectively, greater than 1 GeV/c, 1.5 GeV/c, 1.5 GeV/c, 2 GeV/c, 1.5 GeV/c, 3 GeV/c, 3 GeV/c and 4 GeV/c, which is similar to that observed in minimum-bias pp collisions. The TBW fit parameters and $\chi^2$/NDF for different multiplicity classes in pp collisions at $\sqrt{s}=$ 7 and 13 TeV are, respectively, summarized in tables \ref{tab:parameter_7_TeV_diff_mult} and \ref{tab:parameter_13_TeV_diff_mult}. With these parameters, we present the correlation between $\langle \beta \rangle$ and $T$ in Fig. \ref{fig:beta_T}. As a comparison, the correlation of the parameters in Pb-Pb (Pb-Pb, Xe-Xe, p-Pb) collisions at $\sqrt{s_{\rm NN}}=$ 2.76  (5.02, 5.44, 5.02) TeV is also shown in the figure \cite{TBW_2}. It is found that the evolution of $T$ with $\langle \beta \rangle$ in pp collisions at $\sqrt{s}=$ 13 TeV follows a similar trend as the one observed at $\sqrt{s}=$ 7 TeV, namely $T$ grows with $\langle \beta \rangle$ until it reaches a saturation. Obviously, this trend is opposite to that in pA and AA collisions. Moreover, it is different to the one observed in the BGBW model, in which $T$  non-monotonically depends on $\langle \beta \rangle$ \cite{pi_k_p_vs_mult_pp_13_TeV}. 

\begin{landscape}
\begin{figure}[H]
  \centering
\includegraphics[scale=0.28]{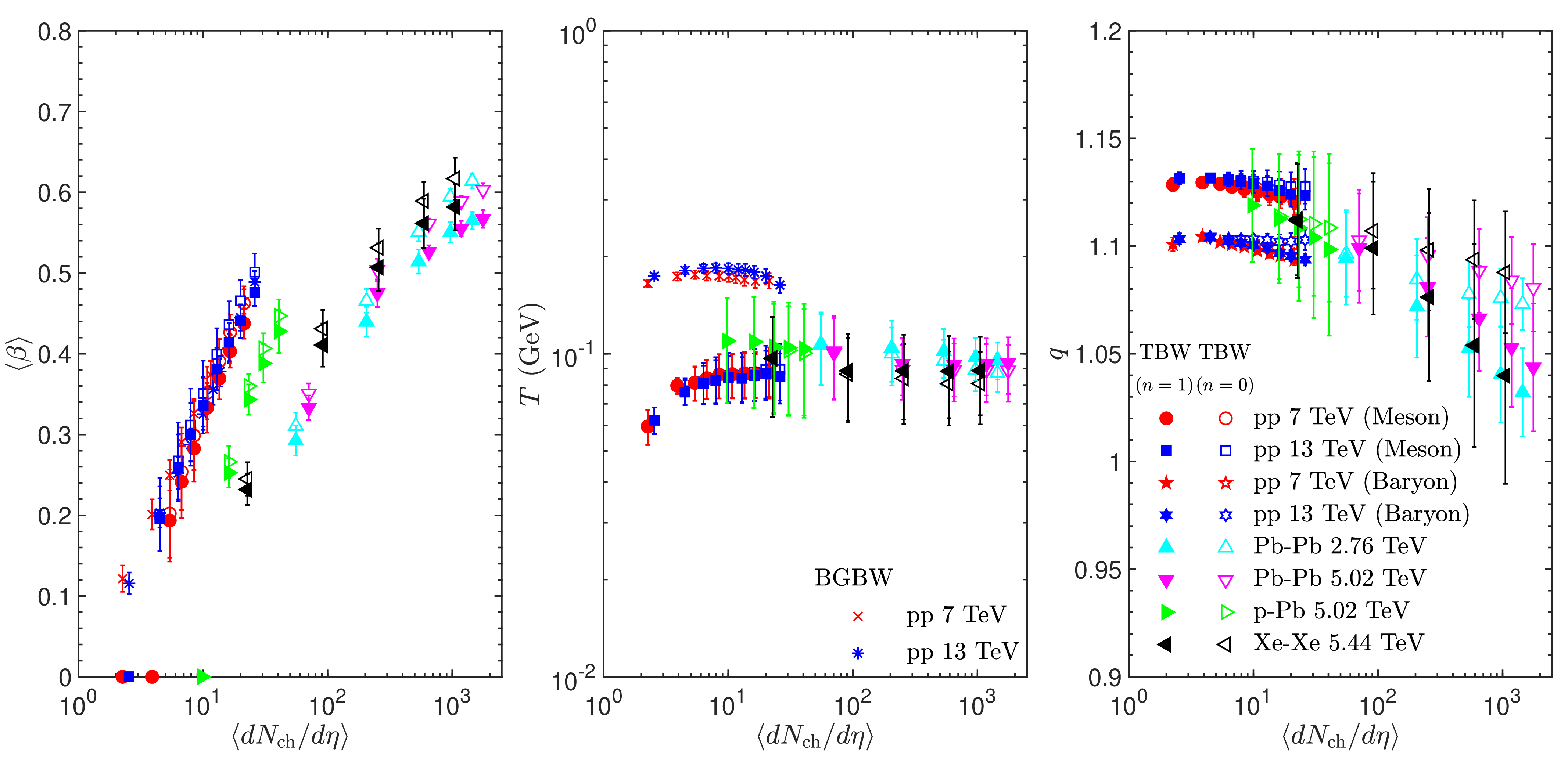}
\caption{\label{fig:beta_T_q_dN_deta} (Colour online) The dependence of $\langle \beta \rangle$, $T$ and $q$ on $\langle dN_{\rm ch}/d\eta \rangle$ for the TBW model in pp (Pb-Pb, Pb-Pb, Xe-Xe, p-Pb) collisions at at $\sqrt{s}=$ 7 and 13 ($\sqrt{s_{\rm NN}}=$ 2.76, 5.02, 5.44, 5.02) TeV. Also shown in the figure are the evolution of $\langle \beta \rangle$ and $T$ with $\langle dN_{\rm ch}/d\eta \rangle$ for the BGBW model in pp collisions at $\sqrt{s}=$ 7 and 13 TeV. The error bars in the left, middle and right panels, respectively, represent the total uncertainties of $\langle \beta \rangle$, $T$ and $q$.}
\end{figure}
\end{landscape}

In addition, the dependence of $\langle \beta \rangle$, $T$ and $q$ on $\langle dN_{\rm ch}/d\eta \rangle$ for the TBW model in different colliding systems  is illustrated in Fig. \ref{fig:beta_T_q_dN_deta}. Also presented in the figure are the evolution of $\langle \beta \rangle$ and $T$ for the BGBW model in pp collisions at $\sqrt{s}=$ 7 and 13 TeV. Several conclusions can be drawn from the figure.

(i) $\langle \beta \rangle$ and $T$ for the TBW model with $n=1$ are, respectively, in agreement with those from the model with $n=0$ within errors. However, at large multiplicities, $q$ for the former is slightly smaller than the latter. This is similar to that observed in central Pb–Pb (Pb-Pb, Xe–Xe, p–Pb) collisions at $\sqrt{s_{\rm NN}}=$ 2.76 (5.02, 5.44, 5.02) TeV\cite{TBW_2}.

(ii) $\langle \beta \rangle$ increases with $\langle dN_{\rm ch}/d\eta \rangle$ and the values of $\langle \beta \rangle$ for pp collisions at different energies are in agreement within error bars, indicating that the small system becomes more explosive at larger multiplicities.  In the common $\langle dN_{\rm ch}/d\eta \rangle$ range,  $\langle \beta \rangle$ is higher in pp collisions than in pA and AA collisions. This could be argued as follows. The initial energy density of the fireball is proportional to the charged-particle multiplicity density, while it is inverse proportional to the transverse overlapping area of the collisions\cite{Bjorken}. At similar multiplicities, the transverse area in pp collisions is smaller than that in pA or AA collisions, resulting a higher initial energy density for the former case. This will make the radial flow stronger in pp collisions than in pA or AA collisions. Therefore, the size of the colliding system has significant effects on the final state particle dynamics\cite{pi_k_p_kstar_phi_vs_mult_pp_7_TeV,size_effect}. 

(iii) The non-extensive parameter $q$ decreases with the increase of $\langle dN_{\rm ch}/d\eta \rangle$, indicating that the system is approaching thermal equilibrium at high multiplicities. This could be explained as follows. At high multiplicities, the energy fluctuation at initial impact is washed out more completely by subsequent hadronic interactions than at low multiplicities. Thus, for the former the effect leaves less footprints on the particle spectra than the latter. In addition, we observe that for pp collisions both $q_M$ and  $q_B$ will reach a saturation at low multiplicities. Moreover, at a given multiplicity, $q_M$ is larger than $q_B$, which is similar to that observed in minimum-bias pp collisions.

(iv) The Tsallis temperature parameter in pA and AA collisions decreases with the increase of $\langle dN_{\rm ch}/d\eta \rangle$. However, in pp collisions it exhibits an opposite behaviour and becomes saturate at high multiplicities, which is similar to the trend observed in the Tsallis parameterization of the hadron $p_{\rm T}$ spectra at different multiplicities in pp collisions at $\sqrt{s}=$ 7 TeV\cite{T_vs_mult}. The contrary behaviour of the Tsallis temperature parameter in pp and AA collisions is not understood yet in the present work. Moreover, we observe that in pp collisions the Tsallis temperature parameter is obviously lower than the kinetic freeze-out temperature in the BGBW model, which is similar to the observation in minimum-bias pp  collisions.

\setcounter{footnote}{0}



\begin{figure}[H]
  \centering
\includegraphics[scale=0.30]{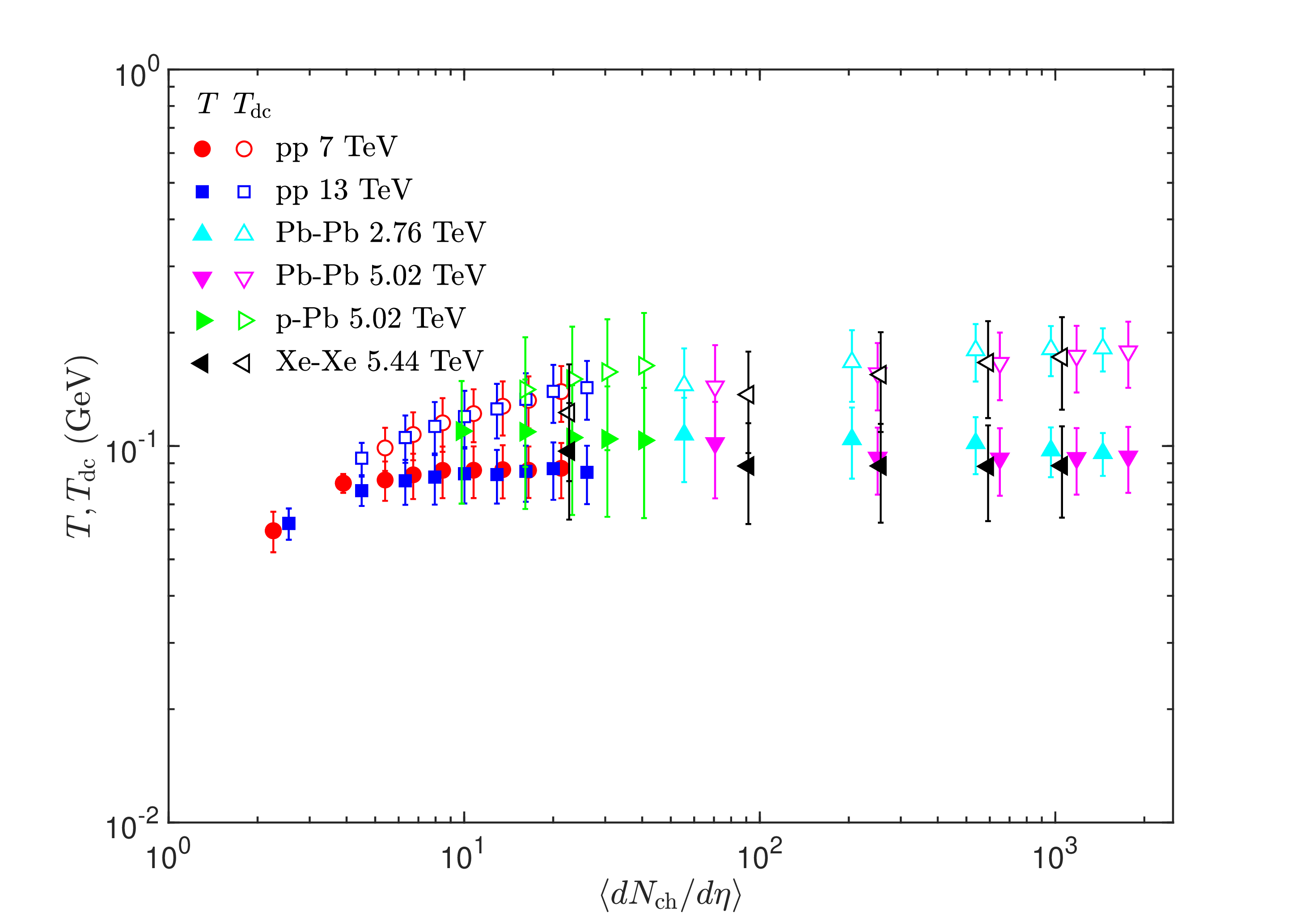}
\caption{\label{fig:T_th_vs_mult} (Colour online) The doppler-corrected temperature parameter as well as the Tsallis temperature parameter for the TBW model with $n=1$ as a function of $\langle dN_{\rm ch}/d\eta \rangle$ for identified hadrons in pp (Pb-Pb, Pb-Pb, Xe-Xe, p-Pb) collisions at $\sqrt{s}=$ 7 and 13 ($\sqrt{s_{\rm NN}}=$ 2.76, 5.02, 5.44, 5.02) TeV. The error bars represent the total uncertainties of the temperature parameter.}
\end{figure}

Finally, as described in ref. \cite{TBW_2}, due to the existence of a radial flow, the doppler-corrected temperature parameter\footnote{It is defined as the inverse slope at high $m_{\rm T}$ or $p_{\rm T}$. The derivation of $T_{\rm dc}$ in the TBW model can be found in the appendix of ref. \cite{TBW_2}.} at the light hadrons freeze-out is larger than the original Tsallis temperature parameter by a blue shift factor, $T_{\rm dc}=T\sqrt{(1+\langle \beta \rangle)/(1-\langle \beta \rangle)}$. This is similar to the case in the BGBW model \cite{BGBW}. Figure \ref{fig:T_th_vs_mult} shows the dependence of the doppler-corrected temperature parameter for the TBW model with $n=1$ in pp, pA and AA collisions. As a comparison, the Tsallis temperature parameter is also presented in the figure. For a given system at a certain collision energy, the doppler-corrected temperature parameter increases with the multiplicity, showing a similar trend as the average transverse momentum \cite{LHC_1,LHC_3,pi_k_p_kstar_phi_vs_mult_pp_7_TeV,pi_k_p_vs_mult_pp_13_TeV,spectra_Xe_Xe_5_44}. Moreover, it is much more strongly multiplicity-dependent than the Tsallis temperature parameter. The reason is obvious. The dependence of the doppler-corrected temperature parameter on the multiplicity originates from two parts: one is the dependence of the Tsallis temperature parameter on $\langle dN_{\rm ch}/d\eta \rangle$; the other is the dependence of the radial flow on $\langle dN_{\rm ch}/d\eta \rangle$. As it can be seen in the left panel of Fig. \ref{fig:beta_T_q_dN_deta}, $\langle \beta \rangle$ increases with $\langle dN_{\rm ch}/d\eta \rangle$ rapidly, which leads to a stronger multiplicity dependence of the doppler-corrected temperature parameter than the Tsallis temperature parameter. Finally, despite the difference in the colliding system and the collision energy, $T_{\rm dc}$ almost scales with the multiplicity in a uniform way, indicating that a universal particle production mechanism may exist in pp, pA and AA collisions at the LHC energies. Indeed, some efforts have been devoted to this universal mechanism, such as the hydrodynamic models\cite{hydro_model_1,hydro_model_2,hydro_model_3}, AMPT\cite{AMPT}, recombination model\cite{RM_1,RM_2}.

\section{\label{sec:conclusions}Conclusions}

In this paper, in order to investigate the evolution of the freeze-out parameters as a function of the collision energy, we have used the Tsallis blast-wave model to simultaneously fit the identified hadron $p_{\rm T}$ spectra in minimum-bias pp collisions at $\sqrt{s}=$ 0.9, 2.76, 5.02, 7 and 13 TeV. In the combined fit,  the Tsallis temperature parameter and the average radial flow velocity are common for all hadrons, while the degrees of non-equilibrium are, respectively, universal for mesons and baryons. We find that the model can describe well the hadron spectra up to 10 GeV/c. The radial flow velocity grows with the increase of the collision energy. The degrees of non-equilibrium and the Tsallis temperature parameter exhibit a similar behaviour as the radial flow velocity, but with a much weaker trend. With the parameterization of the freeze-out parameters on the collision energy, the values of $\langle \beta \rangle$, $T$ and $q$ in pp collisions at $\sqrt{s}=$ 8 and 14 TeV are evaluated and the particle spectra at these two collision energies are predicted.  The agreement between the predicted curves and the experimental data of $\pi^0$, $\eta$, $K^{*0}$ and $\phi$ in pp collisions at $\sqrt{s}=$ 8 TeV implies that our prediction of the $p_{\rm T}$ spectra from the TBW model is reliable. In order to investigate the multiplicity dependence of the freeze-out parameters, the model is extended to the identified particle spectra at different charged-particle multiplicity classes in pp collisions at $\sqrt{s}=$ 7 TeV and 13 TeV. It is observed that at both collision energies the radial flow increases with the multiplicity while the degree of non-equilibrium shows an opposite behaviour, which is similar to that observed in pA and AA collisions at the LHC energies. However, the Tsallis temperature parameter increases with the multiplicity, which is opposite to the trend in pA and AA collisions.  At similar multiplicities, the radial flow in pp collisions is larger than those in pA and AA collisions, indicating that the size of the colliding system has significant effects on the final state particle dynamics.  Finally, an additional flow correction is applied to the Tsallis temperature parameter.  We find that the doppler-corrected temperature parameter depends on the multiplicity in a similar manner,  despite the difference in the colliding system and the collision energy. This indicates that a common particle production mechanism may exist in pp, pA and AA collisions at the LHC energies.

\section*{Acknowledgements}
We wish to thank the ALICE and CMS collaborations for their share of the experimental data. This work is supported by the Fundamental Research Funds for the Central Universities of China under GK202003019, by the Scientific Research Foundation for the Returned Overseas Chinese Scholars, State Education Ministry, by Natural Science Basic Research Plan in Shaanxi Province of China (program No. 2020JM-289) and by the National Natural Science Foundation of China under Grant Nos. 11447024, 11505108, 11905120 and 11947416.

\section*{Appendix}
 \setcounter{equation}{0}
  \setcounter{table}{0}
\setcounter{subsection}{0}
\renewcommand{\theequation}{A\arabic{equation}}
\renewcommand{\thetable}{A\arabic{table}}
\renewcommand{\thesubsection}{A\arabic{subsection}}
The freeze-out parameters in minimum-bias pp collisions and in pp collisions at a given multiplicity class are summarized  in tables \ref{tab:parameter_0_9_13_TeV}, \ref{tab:parameter_7_TeV_diff_mult} and \ref{tab:parameter_13_TeV_diff_mult}. Moreover, we tabulate the values of $\langle dN_{\rm ch}/d\eta \rangle$ at different multiplicity classes in pp collisions and different centralities in pA and AA collisions, respectively, in tables \ref{tab:pp_diff_mult} and  \ref{tab:pA_AA_diff_mult}.

\begin{landscape}
\begin{table}[H]
  \caption{Summary of the freeze-out parameters for the TBW model and the BGBW model in pp collisions at $\sqrt{s}=$ 0.9, 2.76, 5.02, 7 and 13 TeV. The first error is the statistical uncertainty returned from the combined fit, while the second one is the systematic error originating from the variation of the lower bound from 0.5 GeV/c to 0.1 GeV/c for the pion spectrum.}\label{tab:parameter_0_9_13_TeV}
  
	\begin{center}
	\begin{tabular}{c@{\hspace{0.25cm}}c@{\hspace{0.25cm}}c@{\hspace{0.25cm}}c@{\hspace{0.25cm}}c@{\hspace{0.25cm}}c@{\hspace{0.25cm}}c}
		\hline
		
		&$\sqrt{s}$&$\langle \beta \rangle$ & $T$(GeV) & $q_M$ & $q_B$ & $\chi^2$/NDF \\
		\hline
		   &0.9 TeV  &$0.226\pm 0.020 \pm 0.034$ &$0.061\pm 0.003 \pm 0.012$ &$1.1278\pm 0.0024\pm 0.0078$ &$1.0939\pm 0.0026 \pm 0.0029$& 67/94 \\
		 TBW  &2.76 TeV &$0.193\pm 0.012 \pm 0.042$ &$0.086\pm 0.001\pm 0.009$  &$1.1218\pm 0.0005\pm 0.0023$ &$1.0989\pm 0.0007\pm 0.0004$& 105/216 \\
		$(n=1)$&5.02 TeV &$0.249\pm 0.009 \pm 0.031$ &$0.084\pm 0.002 \pm 0.010$ &$1.1260\pm 0.0006 \pm 0.0025$ &$1.1019\pm 0.0007 \pm 0.0010$& 163/231 \\
		   &7 TeV    &$0.262\pm 0.009 \pm 0.042$ &$0.086\pm 0.002 \pm 0.013$ &$1.1272\pm 0.0006 \pm 0.0032$ &$1.1025\pm 0.0007\pm 0.0005$& 191/262 \\
		   &13 TeV   &$0.307\pm 0.007 \pm 0.028$ &$0.085\pm 0.002\pm 0.011$ &$1.1306\pm 0.0006 \pm 0.0029$ &$1.1050\pm 0.0006 \pm 0.0009$& 238/265 \\
		\hline
		   &0.9 TeV  &$0.230\pm 0.022 \pm 0.035$ &$0.062\pm0.003 \pm 0.012$ &$1.1284\pm 0.0024 \pm 0.0078$ &$1.0954\pm 0.0024 \pm 0.0032$& 68/94 \\
		 TBW  &2.76 TeV &$0.200\pm 0.013 \pm 0.045$ &$0.086\pm0.002 \pm 0.010 $ &$1.1220\pm 0.0005 \pm 0.0025$ &$1.0993\pm 0.0006 \pm 0.0008$& 106/216 \\
		$(n=0)$&5.02 TeV &$0.260\pm 0.011 \pm 0.034$ &$0.084\pm 0.002 \pm 0.010$ &$1.1265\pm 0.0006 \pm 0.0027$ &$1.1029\pm 0.0007 \pm 0.0014$& 167/231 \\
		   &7 TeV    &$0.271\pm 0.010 \pm 0.046$ &$0.087\pm 0.002 \pm 0.012$ &$1.1278\pm 0.0006 \pm 0.0036$ &$1.1036\pm 0.0006 \pm 0.0013$& 196/262 \\
		   &13 TeV   &$0.320\pm 0.008 \pm 0.032$ &$0.086\pm 0.002 \pm 0.012$ &$1.1314\pm 0.0006 \pm 0.0034$ &$1.1069\pm 0.0005 \pm 0.0016$& 252/265 \\
		\hline
		 &0.9 TeV  &$0.272\pm 0.016 \pm 0.066$ &$0.142\pm0.004 \pm 0.028$ &--- &---& 51/49 \\
		 &2.76 TeV &$0.246\pm 0.011 \pm 0.107$ &$0.172\pm0.004 \pm 0.053 $ &--- &---& 36/59 \\
       BGBW&5.02 TeV &$0.277\pm 0.010 \pm 0.104$ &$0.171\pm 0.004 \pm 0.055$ &---&---& 29/59 \\
		 &7 TeV    &$0.291\pm 0.011 \pm 0.110$ &$0.174\pm 0.004 \pm 0.059$ &--- &---& 34/59\\
		 &13 TeV   &$0.329\pm 0.013 \pm 0.108$ &$0.174\pm 0.005 \pm 0.059$ &--- &---& 37/54 \\
		\hline
		
	\end{tabular}
         \end{center}
\end{table}
\end{landscape}

\begin{table}[H]
  \caption{The values of $\langle dN_{\rm ch}/d\eta \rangle$ at different multiplicity classes in pp collisions at $\sqrt{s}=$ 7 and 13 TeV. They are taken from refs. \cite{pi_k_p_kstar_phi_vs_mult_pp_7_TeV,pi_k_p_vs_mult_pp_13_TeV}.}\label{tab:pp_diff_mult}
  
  \begin{center}
	\begin{tabular}{c@{\hspace{0.6cm}}c@{\hspace{0.3cm}}c}
		\hline
		
		Class& 7 TeV & 13 TeV \\
		\hline
		I&$21.3 \pm 0.6$&$26.02\pm 0.35$\\
		II&$16.5\pm 0.5$&$20.02\pm 0.27$\\
		III&$13.5\pm 0.4$&$16.17\pm 0.22$\\
		IV+V&$10.76\pm 0.3$&$12.91\pm 0.18$\\
		VI&$8.45\pm 0.25$&$10.02\pm 0.14$\\
		VII&$6.72\pm 0.21$&$7.95\pm 0.11$\\
		VIII&$5.4\pm 0.17$&$6.32\pm 0.09$\\
		IX&$3.9\pm 0.14$&$4.5\pm 0.07$\\
		X&$2.26\pm 0.12$&$2.55\pm 0.04$\\
	
		\hline
	\end{tabular}
         \end{center}
\end{table}

\begin{table}[H]
  \caption{The values of $\langle dN_{\rm ch}/d\eta \rangle$ at different centralities in Pb-Pb (Pb-Pb, Xe-Xe, p-Pb) collisions at $\sqrt{s_{\rm NN}}=$2.76 (5.02, 5.44, 5.02) TeV. They are taken from refs. \cite{LHC_1, LHC_2,LHC_3, LHC_4}.}\label{tab:pA_AA_diff_mult}

  \begin{center}
	\begin{tabular}{c@{\hspace{0.6cm}}c@{\hspace{0.3cm}}c@{\hspace{0.3cm}}c@{\hspace{0.3cm}}c@{\hspace{0.3cm}}c}
		\hline
		Centrality& 0-10\% & 10-20\% & 20-40\% & 40-60\% & 60-80\% \\
		\hline
		Pb-Pb 2.76 TeV &$1447.5 \pm 54.5$&$966\pm 37$&$537.5 \pm 19$&$205 \pm 7.5$&$55.5 \pm 3$\\
		Pb-Pb 5.02 TeV &$1765\pm 51.5$&$1180\pm 31$&$649 \pm 17.5$&$250.5 \pm 10$&$70.6 \pm 4.6$\\
		p-Pb 5.02 TeV &$40.6\pm 0.9$&$30.5\pm 0.7$&$23.2 \pm 0.5$&$16.1 \pm 0.4$&$9.8 \pm 0.2$\\
		\hline
		\hline
		Centrality& 0-10\% & 10-30\% & 30-50\% & 50-70\% & 70-90\% \\
		\hline
		Xe-Xe 5.44 TeV &$1053\pm 25$&$592\pm 14$&$256.5 \pm 6.5$&$91.35 \pm 2.5$&$22.65 \pm 1.1$\\
		\hline
	
	        \end{tabular}
                \end{center}
\end{table}

\begin{landscape}
\begin{table}[H]
	\caption{Summary of the freeze-out parameters for the TBW model at different multiplicity classes  in pp collisions at $\sqrt{s}=$ 7 TeV. The explanation for the quoted errors is the
          same as those in table \ref{tab:parameter_0_9_13_TeV}.}\label{tab:parameter_7_TeV_diff_mult}

        \begin{center}
        
	\begin{tabular}{c@{\hspace{0.25cm}}c@{\hspace{0.25cm}}c@{\hspace{0.25cm}}c@{\hspace{0.25cm}}c@{\hspace{0.25cm}}c@{\hspace{0.25cm}}c}
		
		\hline
		&Class&$\langle \beta \rangle$ & $T$(GeV) & $q_M$ & $q_B$ & $\chi^2$/NDF \\
		\hline
		           &I  &$0437\pm 0.004 \pm 0.018$ &$0.087\pm 0.002 \pm 0.015$ &$1.1206\pm 0.0008\pm 0.0061$ &$1.0937\pm 0.0009 \pm 0.0025$& 112/198 \\
		          &II &$0.403\pm 0.005 \pm 0.020$ &$0.086\pm 0.002\pm 0.013$  &$1.1227\pm 0.0007\pm 0.0052$ &$1.0956\pm 0.0008\pm 0.0021$& 110/198 \\
	                   &III &$0.369\pm 0.005 \pm 0.025$ &$0.087\pm 0.002 \pm 0.014$ &$1.1240\pm 0.0007 \pm 0.0050$ &$1.0965\pm 0.0007 \pm 0.0017$& 112/198 \\
             TBW       &IV+V &$0.333\pm 0.006 \pm 0.030$ &$0.086\pm 0.002 \pm 0.013$ &$1.1251\pm 0.0006 \pm 0.0044$ &$1.0978\pm 0.0007\pm 0.0013$& 105/198 \\
      $(n=1)$         &VI   &$0.283\pm 0.009 \pm 0.040$ &$0.086\pm 0.002\pm 0.013$ &$1.1264\pm 0.0006 \pm 0.0038$ &$1.0997\pm 0.0008 \pm 0.0007$&136/198 \\
		           &VII   &$0.241\pm 0.011 \pm 0.043$ &$0.084\pm 0.002\pm 0.011$ &$1.1275\pm 0.0006 \pm 0.0029$ &$1.1007\pm 0.0008 \pm 0.0001$& 129/198 \\
		           &VIII   &$0.194\pm 0.016 \pm 0.048$ &$0.081\pm 0.002\pm 0.010$ &$1.1288\pm 0.0006 \pm 0.0022$ &$1.1017\pm 0.0009 \pm 0.0003$& 149/197 \\
		           &IX   &0(fixed)                                    &$0.080\pm 0.001\pm 0.004$ &$1.1295\pm 0.0006 \pm 0.0018$ &$1.1044\pm 0.0007 \pm 0.0018$& 166/198 \\
		           &X   &0(fixed)                                    &$0.060\pm 0.002\pm 0.007$ &$1.1285\pm 0.0008 \pm 0.0030$ &$1.1008\pm 0.0009 \pm 0.0031$& 276/197 \\
		\hline
		           &I  &$0.462\pm 0.005 \pm 0.020$ &$0.089\pm 0.002 \pm 0.016$ &$1.1242\pm 0.0007\pm 0.0069$ &$1.1003\pm 0.0007 \pm 0.0041$& 103/198 \\
		          &II &$0.427\pm 0.005 \pm 0.021$ &$0.087\pm 0.001\pm 0.014$  &$1.1254\pm 0.0006\pm 0.0058$ &$1.1004\pm 0.0006\pm 0.0034$& 104/198 \\
	                   &III &$0.391\pm 0.006 \pm 0.027$ &$0.087\pm 0.002 \pm 0.014$ &$1.1259\pm 0.0006 \pm 0.0055$ &$1.1000\pm 0.0006 \pm 0.0029$& 107/198 \\
             TBW       &IV+V &$0.351\pm 0.007 \pm 0.032$ &$0.087\pm 0.002 \pm 0.014$ &$1.1264\pm 0.0006 \pm 0.0049$ &$1.1003\pm 0.0006\pm 0.0024$& 104/198 \\
      $(n=0)$         &VI   &$0.299\pm 0.010 \pm 0.042$ &$0.086\pm 0.002\pm 0.013$ &$1.1272\pm 0.0006 \pm 0.0042$ &$1.1010\pm 0.0007 \pm 0.0016$& 135/198 \\
		           &VII   &$0.254\pm 0.012 \pm 0.046$ &$0.084\pm 0.002\pm 0.012$ &$1.1279\pm 0.0006 \pm 0.0033$ &$1.1015\pm 0.0007 \pm 0.0008$& 130/198 \\
		           &VIII   &$0.202\pm 0.017 \pm 0.052$ &$0.081\pm 0.002\pm 0.010$ &$1.1290\pm 0.0006 \pm 0.0024$ &$1.1021\pm 0.0008 \pm 0.0002$& 150/197 \\
		           &IX   &0(fixed)                                     &$0.080\pm 0.001\pm 0.004$ &$1.1295\pm 0.0006 \pm 0.0018$ &$1.1044\pm 0.0007 \pm 0.0018$& 166/198 \\
		           &X   &0(fixed)                                      &$0.060\pm 0.002\pm 0.007$ &$1.1285\pm 0.0008 \pm 0.0030$ &$1.1008\pm 0.0009 \pm 0.0031$& 276/197 \\
		\hline
		
	\end{tabular}
         \end{center}
\end{table}
\end{landscape}

\begin{landscape}
\begin{table}[H]
  \caption{Summary of the freeze-out parameters for the TBW model at different multiplicity classes  in pp collisions at $\sqrt{s}=$13 TeV. The explanation for the quoted errors is the same as those in table \ref{tab:parameter_0_9_13_TeV}.}\label{tab:parameter_13_TeV_diff_mult}

  \begin{center}
	\begin{tabular}{c@{\hspace{0.25cm}}c@{\hspace{0.25cm}}c@{\hspace{0.25cm}}c@{\hspace{0.25cm}}c@{\hspace{0.25cm}}c@{\hspace{0.25cm}}c}
		\hline
		
		&Class&$\langle \beta \rangle$ & $T$(GeV) & $q_M$ & $q_B$ & $\chi^2$/NDF \\
		\hline
		           &I  &$0.476\pm 0.003 \pm 0.016$ &$0.085\pm 0.002 \pm 0.015$ &$1.1234\pm 0.0008\pm 0.0066$ &$1.0937\pm 0.0008 \pm 0.0026$& 118/200 \\
		          &II &$0.440\pm 0.004 \pm 0.020$ &$0.087\pm 0.001\pm 0.015$  &$1.1240\pm 0.0006\pm 0.0058$ &$1.0954\pm 0.0007\pm 0.0023$& 116/206 \\
	                   &III &$0.414\pm 0.004 \pm 0.023$ &$0.086\pm 0.002 \pm 0.014$ &$1.1258\pm 0.0006 \pm 0.0053$ &$1.0966\pm 0.0007 \pm 0.0019$& 127/206 \\
             TBW       &IV+V &$0.381\pm 0.005 \pm 0.026$ &$0.084\pm 0.002 \pm 0.014$ &$1.1279\pm 0.0006 \pm 0.0046$ &$1.0990\pm 0.0007\pm 0.0015$& 132/206 \\
      $(n=1)$         &VI   &$0.336\pm 0.007 \pm 0.034$ &$0.084\pm 0.002\pm 0.014$ &$1.1289\pm 0.0006 \pm 0.0042$ &$1.1006\pm 0.0007 \pm 0.0011$&165/205 \\
		           &VII   &$0.300\pm 0.009 \pm 0.038$ &$0.083\pm 0.002\pm 0.013$ &$1.1299\pm 0.0007 \pm 0.0035$ &$1.1015\pm 0.0008 \pm 0.0007$& 190/205 \\
		           &VIII   &$0.259\pm 0.012 \pm 0.040$ &$0.081\pm 0.002\pm 0.011$ &$1.1306\pm 0.0007 \pm 0.0028$ &$1.1022\pm 0.0008 \pm 0.0003$& 196/205 \\
		           &IX   &$0.196\pm 0.019 \pm 0.035$  &$0.076\pm 0.002\pm 0.006$ &$1.1315\pm 0.0007 \pm 0.0013$ &$1.1039\pm 0.0010 \pm 0.0002$& 269/205 \\
		           &X   &0(fixed)                                    &$0.062\pm 0.001\pm 0.006$ &$1.1316\pm 0.0008 \pm 0.0026$ &$1.1033\pm 0.0009 \pm 0.0024$& 400/204 \\
		\hline
		           &I  &$0.501\pm 0.005 \pm 0.022$ &$0.089\pm 0.002 \pm 0.018$ &$1.1277\pm 0.0008\pm 0.0080$ &$1.1029\pm 0.0007 \pm 0.0047$& 151/200 \\
		          &II &$0.466\pm 0.005 \pm 0.025$ &$0.089\pm 0.002\pm 0.017$  &$1.1275\pm 0.0007\pm 0.0068$ &$1.1021\pm 0.0006\pm 0.0039$& 134/206 \\
	                   &III &$0.436\pm 0.006 \pm 0.028$ &$0.088\pm 0.002 \pm 0.016$ &$1.1284\pm 0.0007 \pm 0.0062$ &$1.1019\pm 0.0006 \pm 0.0034$& 146/206 \\
             TBW       &IV+V &$0.399\pm 0.006 \pm 0.031$ &$0.086\pm 0.002 \pm 0.015$ &$1.1297\pm 0.0006 \pm 0.0054$ &$1.1029\pm 0.0006\pm 0.0028$& 149/206 \\
      $(n=0)$         &VI   &$0.351\pm 0.008 \pm 0.040$ &$0.086\pm 0.002\pm 0.015$ &$1.1300\pm 0.0007\pm 0.0049$ &$1.1031\pm 0.0007 \pm 0.0022$& 177/205 \\
		           &VII   &$0.312\pm 0.011 \pm 0.044$ &$0.084\pm 0.002\pm 0.014$ &$1.1306\pm 0.0007 \pm 0.0041$ &$1.1032\pm 0.0007 \pm 0.0016$& 199/205 \\
		           &VIII   &$0.267\pm 0.013 \pm 0.046$ &$0.082\pm 0.002\pm 0.012$ &$1.1310\pm 0.0007 \pm 0.0033$ &$1.1033\pm 0.0008 \pm 0.0010$& 202/205 \\
		           &IX   &$0.201\pm 0.021 \pm 0.040$  &$0.077\pm 0.002\pm 0.007$ &$1.1317\pm 0.0007 \pm 0.0015$ &$1.1044\pm 0.0009 \pm 0.0001$& 271/205 \\
		           &X   &0(fixed)                                      &$0.062\pm 0.001\pm 0.006$ &$1.1316\pm 0.0008 \pm 0.0026$ &$1.1033\pm 0.0009 \pm 0.0026$& 400/204 \\
		\hline
	
	\end{tabular}
         \end{center}
\end{table}
\end{landscape}


\section*{References}
\begin{thebibliography}{00}

\bibitem{hydro_modelling} R. de Souza et al. 2016 Prog. Part. Nucl. Phys. \href{http://dx.doi.org/10.1016/j.ppnp.2015.09.002}{ {\bf 86}, 35} (\href{https://arxiv.org/abs/1506.03863}{arXiv:1506.03863})
\bibitem{BGBW} E. Schnedermann, J. Sollfrank, and U. Heinz 1993 Phys. Rev. C \href{http://dx.doi.org/10.1103/PhysRevC.48.2462}{{\bf 48}, 2462} (\href{https://arxiv.org/abs/nucl-th/9307020}{arXiv:nucl-th/9307020})
\bibitem{star_1} J. Adam et al. (STAR Collaboration) 2004 Phys. Rev. Lett.  \href{http://dx.doi.org/10.1103/PhysRevLett.92.112301} {{\bf 92}, 112301}  (\href{https://arxiv.org/abs/nucl-ex/0310004}{arXiv:nucl-ex/0310004})
\bibitem{star_2} B. Abelev et al. (STAR Collaboration) 2009 Phys. Rev. C  \href{http://dx.doi.org/10.1103/PhysRevC.79.034909}{{\bf 79}, 034909}  (\href{https://arxiv.org/abs/0808.2041}{arXiv:0808.2041})
\bibitem{star_3} B. Abelev et al. (STAR Collaboration) 2017 Phys. Rev. C \href{http://dx.doi.org/10.1103/PhysRevC.96.044904}{{\bf 96}, 044904} (\href{https://arxiv.org/abs/1701.07065}{arXiv:1701.07065})
\bibitem{LHC_1} B. Abelev et al. (ALICE Collaboration) 2013 Phys. Rev. C \href{http://dx.doi.org/10.1103/PhysRevC.88.044910} {{\bf 88}, 044910} (\href{https://arxiv.org/abs/1303.0737}{arXiv:1303.0737})
\bibitem{LHC_2} S. Acharya et al. (ALICE Collaboration) 2020 Phys. Rev. C \href{http://dx.doi.org/10.1103/PhysRevC.101.044907 }{{\bf 101}, 044907}  (\href{https://arxiv.org/abs/1910.07678}{arXiv:1910.07678})
\bibitem{LHC_3} B. Abelev et al. (ALICE Collaboration) 2014 Phys. Lett. B  \href{http://dx.doi.org/10.1016/j.physletb.2013.11.020 }{{\bf 728}, 25-38}  (\href{https://arxiv.org/abs/1307.6796}{arXiv:1307.6796})
\bibitem{pi_k_p_kstar_phi_vs_mult_pp_7_TeV} S. Acharya et al. (ALICE Collaboration) 2019 Phys. Rev. C  \href{http://dx.doi.org/10.1103/PhysRevC.99.024906}{{\bf 99}, 024906 } (\href{https://arxiv.org/abs/1807.11321}{arXiv:1807.11321})
\bibitem{pi_k_p_vs_mult_pp_13_TeV} S. Acharya et al. (ALICE Collaboration) 2020 Eur. Phys. J. C \href{http://dx.doi.org/10.1140/epjc/s10052-020-8125-1}{ {\bf 80}, 693} (\href{https://arxiv.org/abs/2003.02394}{arXiv:2003.02394})
\bibitem{pp_double_ridge} G. Aad et al. (ATLAS Collaboration) 2016 Phys. Rev. Lett. \href{http://dx.doi.org/10.1103/PhysRevLett.116.172301}{{\bf 116}, 172301} (\href{https://arxiv.org/abs/1509.04776}{arXiv:1509.04776})
\bibitem{TBW_1} Z. Tang et al. 2009 Phys. Rev. C \href{http://dx.doi.org/10.1103/PhysRevC.79.051901}{{\bf 79}, 051901(R) } (\href{https://arxiv.org/abs/0812.1609}{arXiv:0812.1609})
\bibitem{Tsallis_8} J. Adams et al. (STAR Collaboration) 2006 Phys. Lett. B \href{http://dx.doi.org/10.1016/j.physletb.2006.04.032}{{\bf 637}, 161-169} (\href{https://arxiv.org/abs/nucl-ex/0601033}{arXiv:nucl-ex/0601033})
\bibitem{Tsallis_9} A. Adare et al. (PHENIX Collaboration) 2011 Phys. Rev. C \href{http://dx.doi.org/10.1103/PhysRevC.83.024909 }{{\bf 83}, 024909} (\href{https://arxiv.org/abs/1004.3532}{arXiv:1004.3532})
\bibitem{Tsallis_10} S. Chatrchyan et al. (ALICE Collaboration) 2014 Eur. Phys. J. C \href{http://dx.doi.org/10.1140/epjc/s10052-014-2847-x }{{\bf 74}, 2847}  (\href{https://arxiv.org/abs/1307.3442}{arXiv:1307.3442})
\bibitem{Tsallis_11} M. D. Azmi and J. Cleymans 2014 Acta Phys. Pol. B Proc. Suppl.  \href{http://dx.doi.org/10.5506/APhysPolBSupp.7.9}{ {\bf 7}, 9-16}  (\href{https://arxiv.org/abs/1310.0217}{arXiv:1310.0217})
\bibitem{Tsallis_12} H. Zheng and L. Zhu 2015 Adv. High Energy Phys. \href{http://dx.doi.org/10.1155/2015/180491}{{\bf 2015}, 180491 }(\href{https://arxiv.org/abs/1510.05449}{arXiv:1510.05449})
\bibitem{Tsallis_1} J. Cleymans and D.Worku 2012 J. Phys. G: Nucl. Part. Phys. \href{http://dx.doi.org/10.1088/0954-3899/39/2/025006}{{\bf 39}, 025006} (\href{https://arxiv.org/abs/1110.5526}{arXiv:1110.5526})
\bibitem{Tsallis_2} J. Cleymans et al. 2013 Phys. Lett. B \href{http://dx.doi.org/10.1016/j.physletb.2013.05.029 }{{\bf 723}, 351-354}  (\href{https://arxiv.org/abs/1302.1970}{arXiv:1302.1970})
\bibitem{Tsallis_3} M. D. Azmi and J. Cleymans 2014 J. Phys. G: Nucl. Part. Phys. \href{http://dx.doi.org/10.1088/0954-3899/41/6/065001}{{\bf 41}, 065001} (\href{https://arxiv.org/abs/1401.4835}{arXiv:1401.4835})
\bibitem{Tsallis_4} M. D. Azmi and J. Cleymans 2015 Eur. Phys. J. C \href{http://dx.doi.org/10.1140/epjc/s10052-015-3629-9}{{\bf 75}, 430 }  (\href{https://arxiv.org/abs/1501.07127}{arXiv:1501.07127})
\bibitem{Tsallis_5} L. Marques et al. 2015 Phys. Rev. D \href{http://dx.doi.org/10.1103/PhysRevD.91.054025}{ {\bf 91}, 054025 } (\href{https://arxiv.org/abs/1501.00953}{arXiv:1501.00953})
\bibitem{Tsallis_6} H. Zheng et al. 2015 Phys. Rev. D \href{http://dx.doi.org/10.1103/PhysRevD.92.074009 }{{\bf 92}, 074009} (\href{https://arxiv.org/abs/1506.03156}{arXiv:1506.03156})
\bibitem{Tsallis_7} A. Khuntia et al. 2017 Eur. Phys. J. A \href{http://dx.doi.org/10.1140/epja/i2017-12291-8 }{ {\bf 53}, 103 } (\href{https://arxiv.org/abs/1702.06885}{arXiv:1702.06885})
\bibitem{Tsallis_0} C. Tsallis 1988 J. Stat. Phys. \href{http://dx.doi.org/10.1007/BF01016429}{ {\bf 52}, 479}
\bibitem{TBW_2} G. Che et al. 2021 J. Phys. G: Nucl. Part. Phys. \href{http://dx.doi.org/10.1088/1361-6471/ac09dc}{{\bf 48}, 095103 }  (\href{https://arxiv.org/abs/2010.14880}{arXiv:2010.14880})
\bibitem{TBW_3} J. Chen et al. 2021 Phys. Rev. C  \href{ http://dx.doi.org/10.1103/PhysRevC.104.034901}{{\bf 104}, 034901} (\href{https://arxiv.org/abs/2012.02986}{arXiv:2012.02986})
\bibitem{TBW_4} M. Shao et al. 2010 J. Phys. G: Nucl. Part. Phys.  \href{http://dx.doi.org/10.1088/0954-3899/37/8/085104 }{{\bf 37}, 085104} (\href{https://arxiv.org/abs/0912.0993}{arXiv:0912.0993})
\bibitem{TBW_5} K. Jiang et al. 2015 Phys. Rev. C \href{http://dx.doi.org/10.1103/PhysRevC.91.024910}{{\bf 91}, 024910 }  (\href{https://arxiv.org/abs/1312.4230}{arXiv:1312.4230})
\bibitem{q_meaning}  G. Wilk and Z. Wlodarczyk 2000 Phys. Rev. Lett.  \href{http://dx.doi.org/10.1103/PhysRevLett.84.2770}{{\bf 84}, 2770 } (\href{https://arxiv.org/abs/hep-ph/9908459}{arXiv:hep-ph/9908459})
\bibitem{TBW_v2}  C. Adler et al. (STAR Collaboration) 2001 Phys. Rev. Lett. \href{http://dx.doi.org/10.1103/PhysRevLett.87.182301}{ {\bf 87}, 182301 } (\href{https://arxiv.org/abs/nucl-ex/0107003}{arXiv:nucl-ex/0107003})
\bibitem{pi_k_p_pp_0_9_TeV} K. Aamodt et al. (ALICE Collaboration) 2011 Eur. Phys. J. C \href{http://dx.doi.org/10.1140/epjc/s10052-011-1655-9}{{\bf 71}, 1655}  (\href{https://arxiv.org/abs/1101.4110}{arXiv:1101.4110})
\bibitem{ks_phi_lambda_xi_pp_0_9_TeV} K. Aamodt et al. (ALICE Collaboration) 2011  Eur. Phys. J. C \href{http://dx.doi.org/10.1140/epjc/s10052-011-1594-5 }{{\bf 71}, 1594 }  (\href{https://arxiv.org/abs/1012.3257}{arXiv:1012.3257})
\bibitem{pi_k_p_pp_2_76_TeV} B. Abelev et al. (ALICE Collaboration) 2014 Phys. Lett. B \href{http://dx.doi.org/10.1016/j.physletb.2014.07.011}{ {\bf 736}, 196-207} (\href{https://arxiv.org/abs/1401.1250}{arXiv:1401.1250})
\bibitem{Kstar_phi_pp_2_76_TeV} J. Adam et al. (ALICE Collaboration) 2017 Phys. Rev. C \href{http://dx.doi.org/10.1103/PhysRevC.95.064606}{  {\bf 95}, 064606 }  (\href{https://arxiv.org/abs/1702.00555}{arXiv:1702.00555})
\bibitem{xi_omega_pp_2_76_TeV} D. Colella (for the ALICE Collaboration) 2014 J. Phys. Conf. Ser.  \href{http://dx.doi.org/10.1088/1742-6596/509/1/012090}{ {\bf 509}, 012090} (\href{https://arxiv.org/abs/1311.6003}{arXiv:1311.6003})
\bibitem{lambda_pp_2_76_TeV} L. Hanratty 2014  Ph.D. thesis, Birmingham U.
\bibitem{pi_k_p_pp_5_02_TeV} J. Adam et al. (ALICE Collaboration) 2016 Phys. Lett. B  \href{http://dx.doi.org/10.1016/j.physletb.2016.07.050 }{{\bf 760}, 720-735 } (\href{https://arxiv.org/abs/1601.03658}{arXiv:1601.03658})
\bibitem{kstar_phi_pp_5_02_TeV} S. Acharya et al. (ALICE Collaboration) 2020 Phys. Lett. B  \href{http://dx.doi.org/10.1016/j.physletb.2020.135225 }{ {\bf 802}, 135225} (\href{https://arxiv.org/abs/1910.14419}{arXiv:1910.14419})
\bibitem{ks_lambda_xi_omega_pp_5_02_TeV} A. M. Sirunyan et al. (CMS Collaboration) 2020  Phys. Rev. C \href{http://dx.doi.org/10.1103/PhysRevC.101.064906  }{{\bf 101}, 064906 } (\href{https://arxiv.org/abs/1910.04812}{arXiv:1910.04812})
\bibitem{kstar_phi_pp_7TeV} S. Acharya et al. (ALICE Collaboration) 2020 Phys. Rev. C \href{http://dx.doi.org/10.1103/PhysRevC.102.024912}{ {\bf 102}, 024912 } (\href{https://arxiv.org/abs/1910.14410}{arXiv:1910.14410})
\bibitem{ks_lambda_in_pp_7TeV_and_all_in_pp_13TeV} S. Acharya et al. (ALICE Collaboration) 2021 Eur. Phys. J. C \href{http://dx.doi.org/10.1140/epjc/s10052-020-08690-5}{ {\bf 81}, 256 } (\href{https://arxiv.org/abs/2005.11120}{arXiv:2005.11120})
\bibitem{Omega_Xi_pp_7TeV} B. Abelev et al. (ALICE Collaboration) 2012 Phys. Lett. B  \href{http://dx.doi.org/10.1016/j.physletb.2012.05.011}{{\bf 712}, 309-318 }  (\href{https://arxiv.org/abs/1204.0282}{arXiv:1204.0282})
\bibitem{strange_vs_mult_pp_7TeV} J. Adam et al. (ALICE Collaboration) 2017 Nature Phys. \href{http://dx.doi.org/10.1038/nphys4111}{{\bf 13}, 535} (\href{https://arxiv.org/abs/1606.07424}{arXiv:1606.07424})
\bibitem{kstar_phi_vs_mult_pp_13TeV} S. Acharya et al. (ALICE Collaboration) 2020 Phys. Lett. B \href{http://dx.doi.org/10.1016/j.physletb.2020.135501 }{ {\bf 807}, 135501} (\href{https://arxiv.org/abs/1910.14397}{arXiv:1606.07424})
\bibitem{strange_mult_pp_13_TeV} S. Acharya et al. (ALICE Collaboration) 2020 Eur. Phys. J. C \href{http://dx.doi.org/10.1140/epjc/s10052-020-7673-8 }{{\bf 80}, 167} (\href{https://arxiv.org/abs/1908.01861}{arXiv:1908.01861})
\bibitem{T_dependence_1} H. L. Lao et al. 2017 Eur. Phys. J. A \href{http://dx.doi.org/10.1140/epja/i2017-12238-1}{{\bf 53}, 44 } (\href{https://arxiv.org/abs/1611.08391}{arXiv:1611.08391})
\bibitem{T_dependence_2} H. L. Lao et al. 2018 Nucl. Sci. Tech. \href{http://dx.doi.org/10.1007/s41365-018-0425-x}{{\bf 29}, 82} (\href{https://arxiv.org/abs/1703.04944}{arXiv:1703.04944})
\bibitem{T_dependence_3} S. Zhang et al. 2016 Adv. High Energy Phys. \href{http://dx.doi.org/10.1155/2016/9414239} {{\bf 2016}, 9414239} (\href{https://arxiv.org/abs/1602.01564}{arXiv:1602.01564})
\bibitem{T_dependence_5} B. Abelev et al. (ALICE Collaboration) 2012 Phys. Rev. Lett. \href{http://dx.doi.org/10.1103/PhysRevLett.109.252301} {{\bf 109}, 252301} (\href{https://arxiv.org/abs/1208.1974}{arXiv:1208.1974})
\bibitem{T_dependence_4} A. Andronic 2014 Int. J. Mod. Phys. A \href{http://dx.doi.org/10.1142/S0217751X14300476}{{29}, 1430047} (\href{https://arxiv.org/abs/1407.5003}{arXiv:1407.5003})
\bibitem{power_law_2} C. Y. Wong and G. Wilk 2013 Phys. Rev. D \href{http://dx.doi.org/10.1103/PhysRevD.87.114007}{ {\bf 87}, 114007} (\href{https://arxiv.org/abs/1305.2627}{arXiv:1305.2627})
\bibitem{power_law_3} C. Y. Wong et al. 2015 Phys. Rev. D \href{http://dx.doi.org/10.1103/PhysRevD.91.114027}{ {\bf 91}, 114027} (\href{https://arxiv.org/abs/1505.02022}{arXiv:1505.02022})
\bibitem{power_law_1} P. K. Khandai et al. 2013 Int. J. Mod. Phys. A \href{http://dx.doi.org/10.1142/S0217751X13500668}{ {\bf 28}, 1350066} (\href{https://arxiv.org/abs/1304.6224}{arXiv:1304.6224})
\bibitem{pi0_eta_8_TeV} S. Acharya et al. (ALICE Collaboration) 2018 Eur. Phys. J. C \href{http://dx.doi.org/10.1140/epjc/s10052-018-5612-8}{{\bf 78}, 263} (\href{https://arxiv.org/abs/1708.08745}{arXiv:1708.08745})
\bibitem{Bjorken} J. D. Bjorken 1983 Phys. Rev. D  \href{https://doi.org/10.1103/PhysRevD.27.140}{{\bf 27}, 140}
\bibitem{size_effect} E. Shuryak and I. Zahed 2013 Phys. Rev. C \href{http://dx.doi.org/10.1103/PhysRevC.88.044915}{ {\bf 88}, 044915 } (\href{https://arxiv.org/abs/1301.4470}{arXiv:1301.4470})
\bibitem{T_vs_mult} A. Khuntia et al. 2019 Eur. Phys. J. A \href{https://doi.org/10.1140/epja/i2019-12669-6}{ {\bf 55}, 3} (\href{https://arxiv.org/abs/1808.02383}{arXiv:1808.02383})
\bibitem{spectra_Xe_Xe_5_44} S. Acharya et al. (ALICE Collaboration) 2021 Eur. Phys. J. C \href{http://dx.doi.org/10.1140/epjc/s10052-021-09304-4}{ {\bf 81}, 584 } (\href{https://arxiv.org/abs/2101.03100}{arXiv:2101.03100})
\bibitem{LHC_4} S. Acharya et al. (ALICE Collaboration) 2019 Phys. Lett. B \href{http://dx.doi.org/10.1016/j.physletb.2018.12.048} {{\bf 790}, 35-48 } (\href{https://arxiv.org/abs/1805.04432}{arXiv:1805.04432})
\bibitem{hydro_model_1} W. Zhao et al. 2017 Eur. Phys. J. C \href{https://doi.org/10.1140/epjc/s10052-017-5186-x}{{\bf 77}, 645} (\href{https://arxiv.org/abs/1703.10792}{arXiv:1703.10792})
\bibitem{hydro_model_2} W. Zhao et al. 2020 Phys. Rev. Lett.  \href{https://doi.org/10.1103/PhysRevLett.125.07230}{ {\bf 125}, 072301} (\href{https://arxiv.org/abs/1911.00826}{arXiv:1911.00826})
\bibitem{hydro_model_3} W. Zhao et al.  2018   Phys. Lett. B  \href{https://doi.org/10.1016/j.physletb.2018.03.022}{{\bf 780}, 495-500} (\href{https://arxiv.org/abs/1801.00271}{arXiv:1801.00271})
\bibitem{AMPT} C. Zhang et al. 2021 Phys. Rev. C \href{https://doi.org/10.1103/PhysRevC.104.014908} {{\bf 104}, 014908} (\href{https://arxiv.org/abs/2103.10815}{arXiv:2103.10815})
\bibitem{RM_1} R. C. Hwa and L. Zhu 2018 Phys. Rev. C  \href{https://doi.org/10.1103/PhysRevC.97.054908} {{\bf 97}, 054908}  (\href{https://arxiv.org/abs/1803.08065}{arXiv:1803.08065})
\bibitem{RM_2} L. Zhu, H. Zheng and R. C. Hwa 2021 Phys. Rev. C \href{https://doi.org/10.1103/PhysRevC.104.014902} { {\bf 104}, 014902}  (\href{https://arxiv.org/abs/2105.11161}{arXiv:2105.11161})
\end{thebibliography}
\end{document}